\documentclass[11pt,a4paper]{article}
\usepackage{color,graphics,epsfig,amsmath,amsfonts,citesort}
\pdfoutput=1
\begin{document}

%=======================================================================
% new definitions, abreviations, etc
%=======================================================================

\setlength{\unitlength}{1mm}
\def\noi{\noindent}
\def\lsim{\;\raise0.3ex\hbox{$<$\kern-0.75em\raise-1.1ex\hbox{$\sim$}}\;}
\def\gsim{\;\raise0.3ex\hbox{$>$\kern-0.75em\raise-1.1ex\hbox{$\sim$}}\;}
\def\bce{\begin{center}}
\def\ece{\end{center}}
\def\bea{\begin{eqnarray}}
\def\eea{\end{eqnarray}}
\def\beq{\begin{equation}}
\def\eeq{\end{equation}}
\def\ba{\begin{array}}
\def\ea{\end{array}}
\def\nn{\nonumber}
\def\micro{{\tt micrOMEGAs}}
\def\calc{{\tt CalcHEP}}
\def\comp{{\tt CompHEP}}
\def\nmh{{\sc nmhdecay}}
\def\nmtools{{\tt NMSSMTools}}
\def\susy{{\sc susy}}
\def\wh{\widehat}
\def\wt{\widetilde}
\def\d{\delta}
\def\b{\beta}
\def\D{\Delta}
\def\e{\epsilon}
\def\g{\gamma}
\def\G{\Gamma}
\def\l{\lambda}
\def\k{\kappa}
\def\t{\theta}
\def\s{\sigma}
\def\S{\Sigma}
\def\x{\chi}
\def\sf{\wt f}
\def\bino{{\wt B}}
\def\wino{{\wt W}}
\def\higgsino{{\wt H}}
\def\singlino{{\wt S}}
\def\tb{\tan\!\b}
\def\ctb{\cot\!\b}
\def\cb{\cos\!\b}
\def\sb{\sin\!\b}
\def\sbb{\sin\!2\b}
\def\tw{\tan\!\t_W}
\def\cw{\cos\!\t_W}
\def\sw{\sin\!\t_W}
\def\Oh2{\Omega h^2}
\def\mhf{M_{1/2}}
\def\m{m_0}
\def\A{A_0}
\def\Ak{A_\kappa}
\def\Al{A_\lambda}
\def\ma{m_A}
\def\Msusy{M_{\rm susy}}
\def\ssi{\sigma_{\x p}^{SI}}
\def\ZZ{\hbox {\it Z\hskip -4.pt Z}}
\def\bsg{b\rightarrow s\gamma}
\def\gmu{(g-2)_\mu}
\def\amu{\delta a_\mu}
\def\bsmu{B_s\rightarrow \mu^+\mu^-}
\def\btau{\bar{B}^+\rightarrow \tau^+\nu_\tau}
\renewcommand{\theequation}{\thesection.\arabic{equation}}
\makeatletter
\@addtoreset{equation}{section}
\makeatother

%=======================================================================
% Title Page
%=======================================================================

\begin{flushright}
LAPTH-1287/08\\
LPTA/08-066
\end{flushright}

\bce
{\Large\bf Precision measurements, dark matter direct detection and LHC Higgs searches in a constrained NMSSM} \\[6mm]
{\large G.~B\'elanger$^1$, C.~Hugonie$^2$, A.~Pukhov$^3$} \\[4mm]
{\it\small
1) LAPTH, U. de Savoie, CNRS, B.P.110, F-74941 Annecy-le-Vieux, France \\
2) Laboratoire Physique Th\'eorique et Astroparticules\footnote{Unit\'e Mixte de Recherche -- CNRS -- UMR 5207}\\
Universit\'e de Montpellier II, F-34095 Montpellier, France\\
3) Skobeltsyn Institute of Nuclear Physics,
Moscow State University, 119992 Moscow, Russia
}\\[4mm]
\ece

\begin{abstract}
We reexamine the constrained version of the Next-to-Minimal Supersymmetric Standard Model with semi universal parameters at the GUT scale (CNMSSM). We include constraints from collider searches for Higgs and \susy\ particles, upper bound on the relic density of dark matter, measurements of the muon anomalous magnetic moment and of B-physics observables as well as direct searches for dark matter. We then study the prospects for direct detection of dark matter in large scale detectors and comment on the prospects for discovery of heavy Higgs states at the LHC.
\end{abstract}

%=======================================================================
\section{Introduction}
%=======================================================================

The Next-to-Minimal Supersymmetric Standard Model (NMSSM) is a simple extension of the MSSM that retains two of its most attractive features -- a solution to the hierarchy problem and a natural dark matter (DM) candidate -- while solving the naturalness problem. This is achieved by the introduction of a gauge singlet superfield, $S$. The vev of this singlet field determines the effective $\mu$ parameter which is then naturally of the EW scale~\cite{Frere:1983ag,Nilles:1982dy,Nilles:1982mp,Derendinger:1983bz,Ellis:1988er,Drees:1988fc,Ellwanger:1993xa,Ellwanger:1996gw,King:1995vk,Franke:1995tc}. In the NMSSM the upper bound on the lightest scalar Higgs mass can be larger than in the MSSM~\cite{Yeghian:1999kr,Ellwanger:1999ji,Ellwanger:2005fh,Ellwanger:2006rm,Barger:2006dh} or below the LEP bound if the latter decays into two lighter pseudoscalar Higgs states~\cite{Barger:2006dh,Dobrescu:2000yn,Dobrescu:2000jt,Ellwanger:2005uu,Djouadi:2008uw,Chang:2008cw}. In addition, a very light pseudoscalar Higgs state is not excluded by LEP or B-physics observables~\cite{Dermisek:2006py,SanchisLozano:2007wv,Domingo:2007dx,Domingo:2008rr}. The neutralino sector of the NMSSM contains an extra state, the singlino, which couples only weakly to most of the non-singlet (MSSM like) particles. Owing both to modifications in the Higgs sector and the neutralino sector, the DM properties can differ from those of the MSSM~\cite{Flores:1990bt,Stephan:1997rv,Stephan:1997ds,Belanger:2005kh,Gunion:2005rw,Hugonie:2007vd,Cerdeno:2007sn}. In particular a LSP with a large singlino component has different annihilation properties than the bino LSP that is in general found in the CMSSM. It is even possible to have a very light singlino LSP if accompanied by a very light scalar. Even when the LSP has no singlino component, the more elaborate Higgs sector of the model provides additional channels for rapid annihilation through Higgs exchange. This can have implications for direct and indirect detection rates~\cite{Flores:1991rx,Cerdeno:2004xw,Cerdeno:2007sn,Barger:2007nv,Ferrer:2006hy}.

When imposing GUT scale boundary conditions, the constrained NMSSM (CNMSSM), contains only a handful of parameters, making it convenient for phenomenological analyses. In our definition of the CNMSSM, we impose universal parameters at the GUT scale except in the singlet sector. The model thus contains 6 free parameters plus an arbitrary sign~\cite{Ellwanger:2006rn}. Because of the relations among the various physical parameters at the weak scale, there are stringent constraints on the fundamental parameters of the CNMSSM at the GUT scale. Constraints from Higgs and \susy\ searches at colliders as well as cosmological constraints, resulting from the precise measurement of the relic density of DM were examined in ref.~\cite{Hugonie:2007vd}. Regions of parameter space where the predictions for DM differed from the ones of the CMSSM were highlighted. Apart from the additional allowed parameter space due to relaxed constraints from LEP on the Higgs sector, it was found that rapid annihilation regions (where the LSP annihilates efficiently through a Higgs resonance) could occur at small values of $\tb$, a direct consequence of the more elaborate Higgs sector of the model. However the scenario with a very light singlino LSP annihilating through a very light (and mainly singlet) scalar Higgs state, that can occur in the general NMSSM~\cite{Belanger:2005kh,Gunion:2005rw}, cannot be realised in the CNMSSM.

Within the hypothesis of minimal flavour violation, the B-physics observables put additional constraints on the parameter space of the model~\cite{Hiller:2004ii,Dermisek:2006py,SanchisLozano:2007wv,Domingo:2007dx,Domingo:2008rr}, in particular at large $\tb$ and when the heavy Higgs states are not too heavy. Explaining the discrepancy with the SM expectations for the muon anomalous magnetic moment also restricts the parameter space, in particular sfermions must be light~\cite{Domingo:2008bb,Gunion:2008dg,Domingo:2008rr}. Finally the experiments that have been searching for elastic scattering of DM in large detectors have now reached a sensitivity level where they can constrain the parameter space of \susy\ models including that of the (C)MSSM~\cite{Ahmed:2008eu,deAustri:2006pe} and the (C)NMSSM~\cite{Cerdeno:2004xw,Cerdeno:2007sn,Barger:2007nv}. In particular the LSP with a large higgsino component that is found at large $\m$ and large $\tb$ can lead to cross sections above the present limit. This can also occur in models with a rather light heavy Higgs doublet.

Our first goal in this paper is to reexamine the parameter space of the CNMSSM including in addition to the cosmological constraints already considered in ref.~\cite{Hugonie:2007vd}, the constraints from B-physics observables~\cite{Mack:2007ra,Barberio:2008fa,Domingo:2007dx}, from the muon anomalous magnetic moment~\cite{Zhang:2008pk,Domingo:2008bb} and from the recent upper limit on DM direct detection~\cite{Angle:2007uj,Ahmed:2008eu}. After having identified the allowed regions we concentrate on those which differ markedly from the MSSM, where either the LSP has some singlino component or there is a possibility of LSP annihilation through a (singlet) Higgs resonance. We then explore the potential of future direct detection experiments for probing the remaining parameter space of the model. We find that with the exception of the singlino LSP scenarios, large scale detectors will probe most of the parameter space of the model. Signals are in several cases expected just beyond the present limits.

As a secondary goal of this paper we address the issue of distinguishing the CNMSSM from the CMSSM in regions favoured by the cosmological measurements. Assuming that a signal compatible with \susy\ is observed in flavour physics, DM detection and/or at the LHC, we investigate the LHC potential for probing the NMSSM Higgs sector. The Higgs sector of the NMSSM includes, as in the MSSM, a light scalar state $h$, two almost degenerate heavy scalar/pseudoscalar states $H/A$, and acharged state $H^\pm$. In addition, the NMSSM Higgs sector contains two extra singlet states, a scalar $s$ and a pseudoscalar $a$, whose mass can be below or above that of the (MSSM like) doublet states. A clear indication in favour of the NMSSM would therefore be the discovery of at least three neutral states with large mass splittings. The difficulty in observing the extra singlet states, $s$ or $a$, is a consequence of their tiny couplings to SM particles unless their mixing with the doublet states are substantial. The LHC potential for probing the Higgs sector of singlet extensions of the MSSM, including the NMSSM, was explored in~\cite{Dobrescu:2000jt,Ellwanger:2005uu,Moretti:2006hq,Barger:2006sk,Cheung:2007sva,Forshaw:2007ra,Carena:2007jk,Belyaev:2008gj,Djouadi:2008uw} with emphasis on the light Higgs doublet/singlet. The production of the heavy states  was not considered. The possibility of distinguishing the NMSSM from the MSSM by observing a singlino LSP will not be discussed here~\cite{Ellwanger:1997jj,MoortgatPick:2005vs,Kraml:2005nx,Barger:2006kt,LH:2008gva,Kraml:2008zr}.

The paper is organized as follows: in sec.~\ref{sec:model} we briefly describe the model. In sec.~\ref{sec:constraints} we present the various constraints on the model. Typical case studies are presented in sec.~\ref{sec:res} The direct detection prospects are discussed in sec.~\ref{sec:dd} followed by Higgs searches at the LHC in sec.~\ref{sec:Higgs}. Our results are summarized in sec.~\ref{sec:dis}.

%=======================================================================
\section{The CNMSSM}\label{sec:model}
%=======================================================================

We discuss here the NMSSM with the scale invariant superpotential
\beq
W = \l {S} {H}_u {H}_d + \frac{\k}{3} \, {S}^3 + ({\rm Yukawa\ couplings}) \, ,
\eeq
where no supersymmetric dimensionful parameters as $\mu$ are present in the superpotential, i.e. the weak scale originates from the soft \susy\ breaking scale {\it only}. The soft \susy\ breaking terms in the Higgs sector are then given by
\begin{eqnarray}
V_{\rm soft} & = & m_{H_u}^2 |H_u|^2 + m_{H_d}^2 |H_d|^2 + m_S^2 |S|^2 + \left( \l \Al H_u H_d S + \frac{1}{3} \k \Ak S^3 + \rm{h.c.} \right) \, .
\end{eqnarray}
A possible cosmological domain wall problem caused by a global $\ZZ_3$ symmetry~\cite{Abel:1995wk} can be avoided by introducing suitable non-renormalizable operators~\cite{Abel:1996cr,Panagiotakopoulos:1998yw} which neither generate dangerously large singlet tadpole diagrams~\cite{Nilles:1982mp,Ellwanger:1983mg,Bagger:1993ji,Bagger:1995ay} nor affect the low energy phenomenology. Other models solving the MSSM $\mu$ problem with extra gauge singlet fields and with no domain walls include the nearly minimal \susy\ model (nMSSM), with an additional tadpole term for the singlet in the superpotential and/or in the soft scalar potential~\cite{Panagiotakopoulos:1999ah,Panagiotakopoulos:2000wp,Dedes:2000jp}, and the UMSSM, with an extra $U(1)'$ gauge
symmetry~\cite{Suematsu:1994qm,Cvetic:1997ky,Demir:2005ti,King:2005jy,Barger:2006dh,Choi:2006fz}. The DM properties as well as the possibility of generating the baryon asymmetry of the universe within these models have been analysed in refs.~\cite{Hugonie:2003yu,Menon:2004wv,Huber:2006wf,Balazs:2007pf,deCarlos:1997yv,Suematsu:2005bc,Nakamura:2006ht,Kang:2004pp,Barger:2005hb,Barger:2007nv,Kalinowski:2008iq}.

We consider here a constrained version of the NMSSM (CNMSSM) with semi-universal parameters defined at the GUT scale. Embedding the (N)MSSM in an underlying theory and assuming some spontaneous symmetry breaking mechanism (for example supergravity) typically leads to simple patterns for soft \susy\ breaking parameters at the GUT scale, hence reducing significantly the number of arbitrary parameters in the model. The choice of semi-universality is motivated by the fact than imposing strict universality makes it difficult to find a consistent set of parameters in the NMSSM~\cite{Djouadi:2008yj}. The reason is the following: in the CMSSM, in addition to the soft terms $\A,\m,\mhf$ defined at the GUT scale, $M_Z$ and $\tb$ at the weak scale are used as inputs. The two minimization equations of the Higgs potential w.r.t.\ the two real Higgs vevs $h_u$ and $h_d$ are used to compute $\mu$ and $B$ in terms of the other parameters (this leaves the sign of $\mu$ as a free parameter). Both $\mu$ and $B$ have only a small effect on the renormalization group equations (RGEs) of the other parameters (in numerical codes, this is usually solved by an iterative procedure). On the other hand, in the CNMSSM neither $\mu$ nor $B$ are present and one has to cope with three coupled minimization equations w.r.t.\ the Higgs vevs $h_u$, $h_d$ and $s$. This means that starting from strict universality, with $\mhf$, $\m$, $\A$ as well as $\l$ and $\k$ as free parameters at the GUT scale, one usually ends up with the wrong value of $M_Z$ at the weak scale (as no dimensionful parameter is left to tune the correct value). In addition, $\tb$ cannot be a free parameter in this approach.

The semi-universal approach described here was first used in the program NMSPEC~\cite{Ellwanger:2006rn}: First, let us assume that $\l$ as well as all the soft terms (except $m_S^2$) are known at the weak scale (e.g. after integration of the RGEs down from the GUT scale,). One can define effective ($s$ dependent) parameters at the weak scale:
\beq\label{eq:def}
\mu = \l s \, , \qquad \nu = \k s \, , \qquad B = \Al + \nu \, .
\eeq
The minimization equations w.r.t.\ $h_u$ and $h_d$ can then be solved for the effective $\mu$ and $B$, as in the MSSM, in terms of the other parameters (incl. $M_Z$ and $\tb$). From $\mu$ and $B$ one then deduces (for $\l$ and $A_\l$ given) both $s$ and $\k$. Finally, from the minimization equation w.r.t.\ $s$, one can easily obtain the soft singlet mass $m_S^2$ in terms of all other parameters. At tree level, the minimization equations giving $\mu$ (up to a sign), $B$ and $m_S^2$ read:
\begin{eqnarray}\label{eq:min}
\mu^2 & = & \frac{m_{H_d}^2 - m_{H_u}^2 \tan^2\!\b}{\tan^2\!\b - 1} - \frac{1}{2}M_Z^2 \, , \nn \\
B & = & \frac{\sbb}{2\mu} \left( m_{H_u}^2 + m_{H_d}^2 + 2\mu^2 + \l^2(h_u^2+h_d^2) \right) \, , \\
m_S^2 & = & \l^2\frac{h_u h_d}{\mu}(\Al + 2\nu) - \nu(\Ak + 2\nu) - \l^2(h_u^2+h_d^2) \, . \nn
\end{eqnarray}
The radiative corrections to the scalar potential show a weak dependence on $s$, $\k$ and $m_S^2$ which can be included in the minimization equations. These become non-linear in the parameters to solve for and have therefore to be solved iteratively. The derived parameters $\k$ and $m_S^2$ affect the RGEs of the other parameters not only through threshold effects around $\Msusy$, but also through the $\beta$ functions. However, the numerical impact is relatively small such that an iterative procedure converges quite rapidly again.

As $m_S^2$ is computed from the minimization equations, it is difficult to find parameters such that it assumes the same value as the Higgs doublet (or other scalar) soft masses squared at the GUT scale. On the other hand, the mechanism for the generation of soft \susy\ breaking terms could easily treat the singlet differently from the other non-singlet matter multiplets~\cite{Brax:1994ae}. This would also affect the coupling $\Ak$ which involves the singlet only. Hence, in our semi-universal approach, $\k$, $s$ and $m_S^2$ are computed from the minimization equations leaving the following free parameters:\\[1mm]
\indent $\bullet$ $\tb$, sign($\mu$) at the weak scale;\\[1mm]
\indent $\bullet$ $\l$ at the \susy\ scale;\\[1mm]
\indent $\bullet$ $\mhf$, $\m$, $\A$ and $\Ak$ at the GUT scale.\\[1mm]
Not all choices of parameters are allowed in the CNMSSM. Some lead to negative mass squared for scalar fields (Higgs or sfermions), others to Landau poles below the GUT scale for the dimensionless couplings. The Landau pole condition leads to an upper bound on $\l\!\lsim\! .7$ which depends on $\tb$. The LEP bounds on the Higgs mass further constrain values of $\l\!\gsim\! .1$. This is due to the the fact that although the diagonal entry for the light double $h$ in the scalar Higgs mass matrix increases with $\l$:
\beq\label{eq:mh}
m_h^2 = \left(\cos^2\! 2\b + \displaystyle\frac{\l^2}{g^2} \sin^2\! 2\b \right) M_Z^2 \, ,
\eeq
the mixing between the doublet and the singlet states is proportional to $\l$ times some combination of soft \susy\ breaking terms. In the general NMSSM, this combination can be made to vanish. In the semi-universal NMSSM however, this is no more the case. Therefore, if $\l$ is too large ($\l > .1$) mixing effects lead to a light eigenstate in the Higgs spectrum with a substantial doublet component. Such a state is usually excluded by LEP, unless it decays to two lighter pseudoscalar singlet Higgs states. Such a situation requires however some fine tuning, especially on $\Ak$, and will not be studied here.

When $\l$ is small the mixings between the Higgs doublet and singlet states are small. As in the MSSM, the masses of the nearly degenerate heavy Higgs doublet states read
\beq\label{eq:doublets}
m_H^2 \sim m_A^2 = \frac{2 \mu B}{\sbb} \, .
\eeq
We will denote the scalar singlet $s$ and the pseudoscalar singlet $a$ and their masses read, respectively:
\beq\label{eq:singlets}
m_s^2 = \nu (\Ak+4\nu) \, , \quad m_a^2 = -3 \nu \Ak \, .
\eeq
The parameter $\Ak$ being only slightly renormalized from the GUT scale down to the \susy\ scale, eq.~(\ref{eq:singlets}) shows that the masses of the singlet states are proportional to the value of $\Ak$ at the GUT scale. The condition that both squared masses are positive together with eq.~(\ref{eq:def}) implies
\beq\label{eq:rel}
-4(B - \Al)^2\ \lsim\ \Ak (B - \Al) \ \lsim\ 0 \, .
\eeq
The parameter $B$, obtained from the minimization equations~(\ref{eq:min}) depends on $\m$, $\mhf$, $\l$ and $\tb$, while $\Al$ depends on $\A$, $\l$ and $\tb$ through the RGEs. This means that, for sign($\mu$) positive (which we will always assume in the following), either $\Ak > 0$ and $\A > \wt \A(\m,\mhf,\l,\tb)$ or $\Ak < 0$ and $\A < \wt \A(\m,\mhf,\l,\tb)$. Moreover, for $\Ak > 0$ and $\tb$ moderate, large values of $\m$ or $\mhf$ (implying $B$ large and positive) lead to a negative mass squared for the pseudoscalar singlet $a$ and are therefore disallowed. If $\Ak > 0$ and $\tb$ is large, however, $B$ remains small up to large values for $\m$ and $\mhf$ which are no longer excluded.

%=======================================================================
\section{Constraints}\label{sec:constraints}
%=======================================================================

To evaluate numerically the supersymmetric spectrum we used the NMSPEC program from the \nmtools\ package~\cite{Ellwanger:2006rn}. Starting from GUT scale parameters and using the RGEs, this code computes the Higgs spectrum including higher order corrections~\cite{Ellwanger:1999ji,Ellwanger:2005fh} as well as the masses of sparticles at one-loop. The \nmtools\ package also includes all the available experimental constraints from sparticle and Higgs searches at LEP and Tevatron (for details on the exclusion channels see refs.~\cite{Ellwanger:2004xm,Ellwanger:2005dv,Ellwanger:2006rn}) as well as constraints from B-physics~\cite{Domingo:2007dx} and from the muon anomalous magnetic moment~\cite{Domingo:2008bb}. For the computation of the relic density of DM we rely on \micro~\cite{Belanger:2006is} which is included in the \nmtools\ package. The elastic scattering neutralino nucleon cross section is computed with \micro2.2~\cite{Belanger:2008sj}. We used the central value of the top quark mass measured at the Tevatron, $m_t = 172.6$~GeV, and chose $m_b(m_b) = 4.214$~GeV and $\alpha_S(M_Z) = .1172$. Note that the top quark mass does affect the value of the light Higgs mass as well as the calculation of the spectrum, especially at large $\m$. As stated above, we assumed sign($\mu$) $> 0$.

%%%%%%%%%%%%%%%%%%%%%%%%%%%%%%%%%%%%%
\subsection{B-physics}
%%%%%%%%%%%%%%%%%%%%%%%%%%%%%%%%%%%%%

Assuming minimal flavour violation, a severe constraint on the CNMSSM originates from the branching ratio $Br(\bsg)$. We require that the theoretical prediction, including uncertainties, falls within the $2\sigma$ range~\cite{Barberio:2008fa}:
\beq
3.07\times 10^{-4} <Br(\bsg) < 4.07\times 10^{-4} \, .
\eeq
The most important \susy\ contributions arises from chargino-squarks loops and charged Higgs-top quark loops. For $\mu>0$ considered here, the chargino exchange diagram gives a negative contribution relative to the SM one so that the branching ratio for $\bsg$ drops below the allowed range when the chargino is light, that is in the low $\m$, $\mhf$ region. As in the CMSSM there is a strong dependence on $A_t$ (hence $\A$) from the mixing in the stop sector. The chargino and charged Higgs contribution partially cancel each other when $A_t>0$ thus the $\bsg$ constraint is most effective for negative values of $A_t$ (or $\A$). Also as in the CMSSM, this constraint is stronger for large values of $\tb$.

The neutral Higgs contribution to the branching ratios for $\bsmu$ and the charged Higgs contribution to $\btau$ both feature a strong $\tb$-enhancement (respectively $\propto \tan^6\!\beta$ and $\tan^4\!\beta$). The latest $95\%C.L.$ bound from CDF at Tevatron~\cite{Mack:2007ra},
\beq
Br(\bsmu) < 5.8\times 10^{-8}
\eeq
and the $2\sigma$ world average~\cite{Barberio:2008fa}
\beq
0.34\times10^{-4}<Br(\btau)<2.30\times 10^{-4}
\eeq
thus constrain the large $\tb$ regions where the heavy neutral/charged Higgs are rather light. These conditions are met at low values of $\m$ as well as in the so-called `focus point' region at large $\m$. The relative importance of each channel is influenced by the trilinear coupling $A_t$ which control the flavour violating neutral Higgs coupling. In particular for $\bsmu$ the chargino/stop contribution is suppressed for small values of $A_t$ (obtained when $\A$ is negative). On the other hand $\btau$ is independent of the trilinear coupling while negative values of $\A$ strenghten the constraints from $\bsg$.

The constraints from $\Delta M_s,\Delta M_d$ are also taken into account in \nmtools~\cite{Domingo:2007dx}. Given the present experimental results these observables do not allow to constrain additional parameter space relative to other B-physics observables.

%%%%%%%%%%%%%%%%%%%%%%%%%%%%%%%%%%%%%
\subsection{Muon anomalous magnetic moment}
%%%%%%%%%%%%%%%%%%%%%%%%%%%%%%%%%%%%%

The anomalous magnetic moment of the muon shows a deviation from the SM prediction when using only $e^+e^-$ data to estimate the leading order hadronic SM contribution~\cite{Zhang:2008pk}. We imposed the $2\sigma$ bound:
\beq
8.7\times 10^{-10} < \amu < 4.6\times 10^{-9} \, .
\eeq
As in the CMSSM, the most important supersymmetric contribution comes from the chargino/sneu\-trino exchange and is proportionnal to the Yukawa coupling (which means $\propto \tb$). To explain the deviation from the SM prediction this observable therefore requires light charginos and smuons, hence favours the low $\m$, $\mhf$ region. At larger values of $\tb$ the allowed band moves towards larger values of $\m$, $\mhf$ since light sparticles can then give a contribution that is too large~\cite{Domingo:2008bb}.

%%%%%%%%%%%%%%%%%%%%%%%%%%%%%%%%%%%%%
\subsection{Relic density}
%%%%%%%%%%%%%%%%%%%%%%%%%%%%%%%%%%%%%

The allowed range for the DM relic density from cosmological data~\cite{Spergel:2006hy,Tegmark:2006az} is
\beq
.094<\Omega h^2< .136
\eeq
when including some uncertainty on the cosmological model parameters~\cite{Hamann:2006pf}. We only use the upper bound on $\Omega h^2$ since we assume that the neutralino does not necessarily account for all the DM.

A complete calculation of the neutralino DM in the (C)NMSSM~\cite{Gunion:2005rw,Belanger:2005kh,Hugonie:2007vd,Cerdeno:2007sn} has shown that often the same mechanisms as in the MSSM for neutralino annihilation are at work. Nevertheless important differences are found: the extra pseudoscalar singlet Higgs state $a$ opens up the possibility of resonant annihilation and the singlino component of the neutralino LSP can alter the prediction for annihilation cross sections. The new mechanisms that can provide a DM candidate compatible with the WMAP results are:\\[1mm]
\indent $\bullet$ Annihilation near the pseudoscalar singlet $a$ resonance, occurring for $\l \sim .1$.\\[1mm]
\indent $\bullet$ Coannihilation of the singlino LSP with a stau or stop NLSP for $\l\!\ll\! 1$ and small $\m$.\\[1mm]
\indent $\bullet$ Coannihilation of the singlino LSP with a higgsino NLSP for $\l\!\ll\! 1$, large $\m$ and $\tb$. \\[1mm]
\indent $\bullet$ Coannihilation of the singlino LSP with the bino NLSP, the bino rapidly annihilating \par through a Higgs resonance, for $\l\!\ll\! 1$ and large $\tb$.

%%%%%%%%%%%%%%%%%%%%%%%%%%%%%%%%%%%%%
\subsection{Direct detection}
%%%%%%%%%%%%%%%%%%%%%%%%%%%%%%%%%%%%%

The best limits on the spin independent (SI) neutralino-proton cross section have been set recently by Xenon~\cite{Angle:2007uj} and CDMS~\cite{Ahmed:2008eu}. We will use the mass dependent CDMS limit since this detector has the best sensitivity for masses above $\sim 50$~GeV. In our model a lighter neutralino is excluded by the LEP limit on the chargino mass because of the universality condition on the gaugino masses. The CDMS limit corresponds to $\sigma_p^{SI}> 4.6\times 10^{-8}$~pb for $m_\x=60$~GeV and $\sigma_p^{SI}> 2.5\times 10^{-7}$~pb for $m_\x=670$~GeV. Those limits are obtained by making standard assumptions about the DM density and velocity distribution as well as for specific choices of nuclear form factors. The theoretical uncertainties associated with the choice of the velocity distribution have been estimated to be about a factor 3~\cite{Bottino:2005qj}. These uncertainties will not explicitly be taken into account here. When deriving constraints on the parameter space additional uncertainties that arise from our lack of precise knowledge of the quark coefficents in the nucleon are taken into account by choosing two different sets of coefficients: the default values of \micro~\cite{Belanger:2008sj} corresponding to $(\sigma_{\pi N}, \sigma_0) = (55, 35)$~MeV and $(45, 40)$~MeV. The latter implies a lower $s$ quark coefficient in the nucleon. First lattice calculations have recently obtained results that favour a not too large value for $\sigma_{\pi N}$~\cite{Ohki:2008ff}.

The SI interaction is usually dominated by Higgs exchange provided the neutralino LSP has enough higgsino component to couple to a Higgs. The light Higgs doublet often gives the largest contribution although the contribution of the heavy Higgs doublet can dominate at large $\tb$ when its couplings to neutralinos and quarks are enhanced. As in the MSSM we therefore expect large cross sections when the heavy Higgs is rather light. This occurs both at low $\m$ and in the focus point region at large $\m$. In the latter case the elastic cross section should be specially enhanced at large $\tb$ due both to increased Higgs couplings and to a LSP with a large higgsino fraction. The squark exchange diagram also contributes but is suppressed by the large squark mass except in the low $\m$, $\mhf$ region. This contribution is however never large enough by itself to obtain $\ssi$ near the experimental bound. Note that a pure singlino LSP has no coupling to the Higgs doublets nor to the squarks so one expects an extremely small SI cross section in this scenario.

A higgsino LSP also couples to the $Z$ boson and thus contributes to the spin dependent (SD) cross section. The experimental limits ($\sigma_p^{SD}\approx 10^{-2}$~pb~\cite{Angle:2008we,Lee.:2007qn}) are still orders of magnitude too large for this process to put constraints on the model. We will not consider it in the following.

%=======================================================================
\section{Results}\label{sec:res}
%=======================================================================

We first performed a general scan over the parameter space of the CNMSSM for fixed values of the SM parameters in order to find regions satisfying all theoretical and experimental constraints. We did not find any scenarios satisfying all constraints for small values of $\tb \lsim 5$. This is mainly a result of the LEP constraints on the lightest Higgs boson coupled with the $\gmu$ constraint. In particular, the allowed regions at small values of $\tb = 2$ analyzed in ref.~\cite{Hugonie:2007vd} cannot explain the discrepancy in the measurement of the $\gmu$ (this constraint was not included in this former analysis). We then picked fixed values for $\l$ and $\tb$: $\l = .1$ or $.01$, $\tb = 5, 10$ or $50$ and for each choice scanned randomly on the remaining free parameters, namely $\m$, $\mhf$, $\A$ and $\Ak$. As in our previous analysis, the regions in the parameter space allowed by all theoretical and experimental constraints correspond to either $\Ak < 0$ and $\A < \wt \A(\l,\tb)$ or $\Ak > 0$ and $\A > \wt \A(\l,\tb)$, the latter case appearing only if $\l = .01$ and/or $\tb = 50$ (the other values of $\l$, $\tb$ always lead to light states in the Higgs sector excluded by LEP when $\Ak > 0$).

Finally, we identified the values of $\A$ and $\Ak$ for which the main neutralino annihilation channel is the pseudoscalar singlet $a$ resonance, or the LSP is mainly singlino. This will be illustrated in the following when we present plots in the $\m,\mhf$ plane for selected values of $\l$, $\tb$, $\A$ and $\Ak$.

%%%%%%%%%%%%%%%%%%%%%%%%%%%%%%%%%%%%%
\subsection{Large $\l$: singlet resonances}\label{sec:pseudo}
%%%%%%%%%%%%%%%%%%%%%%%%%%%%%%%%%%%%%

When $\l$ is large ($\gsim .1$), the singlino component of the LSP is always small. Scenarios characteristic of the CNMSSM are therefore only those where the bino LSP annihilation proceeds through the pseudoscalar singlet $a$ resonance. As explained in sec.~\ref{sec:model}, the larger $\l$, the stronger the constraints from Higgs searches at LEP. These can still be evaded if $\Ak$ is such that the light Higgs doublet $h$ decays mainly into two light pseudoscalar singlets, $h \to aa$~\cite{Djouadi:2008uw}. This requires however some fine-tuning on $\Ak$. Furthermore, the $a$ being very light, it cannot be used as a resonance in order to obtain the correct value for the DM relic density. In such case, one has to rely on the coannihilation of the bino LSP with light sfermions (occurring at small $\m$) in order to satisfy the WMAP constraints, as in the CMSSM~\cite{Djouadi:2008uw}. We will not discuss this case here. Another possibility is to maximize the light Higgs doublet mass by taking a small value of $\tb$ (e.g. for $\tb=2$ and $\l=.5$, as in ref.~\cite{Hugonie:2007vd}). However, these regions are incompatible with the $\gmu$ constraint as stated above. We therefore consider only $\l = .1$ and $\tb =5$, $10$ and $50$. In these cases, LEP constraints on the light Higgs doublet imply small $\Ak < 0$ and large $\A < 0$, except for $\tb = 50$ where one can also have small $\Ak > 0$ and large $\A > 0$.

%%%%%%%%%%%%%%%%%%%%%%%%%%%%%%%%%%%%%
\begin{figure}[t!]
\vspace*{-5mm}
\hspace*{-5mm}
\includegraphics[width=.5\textwidth,keepaspectratio,angle=90]{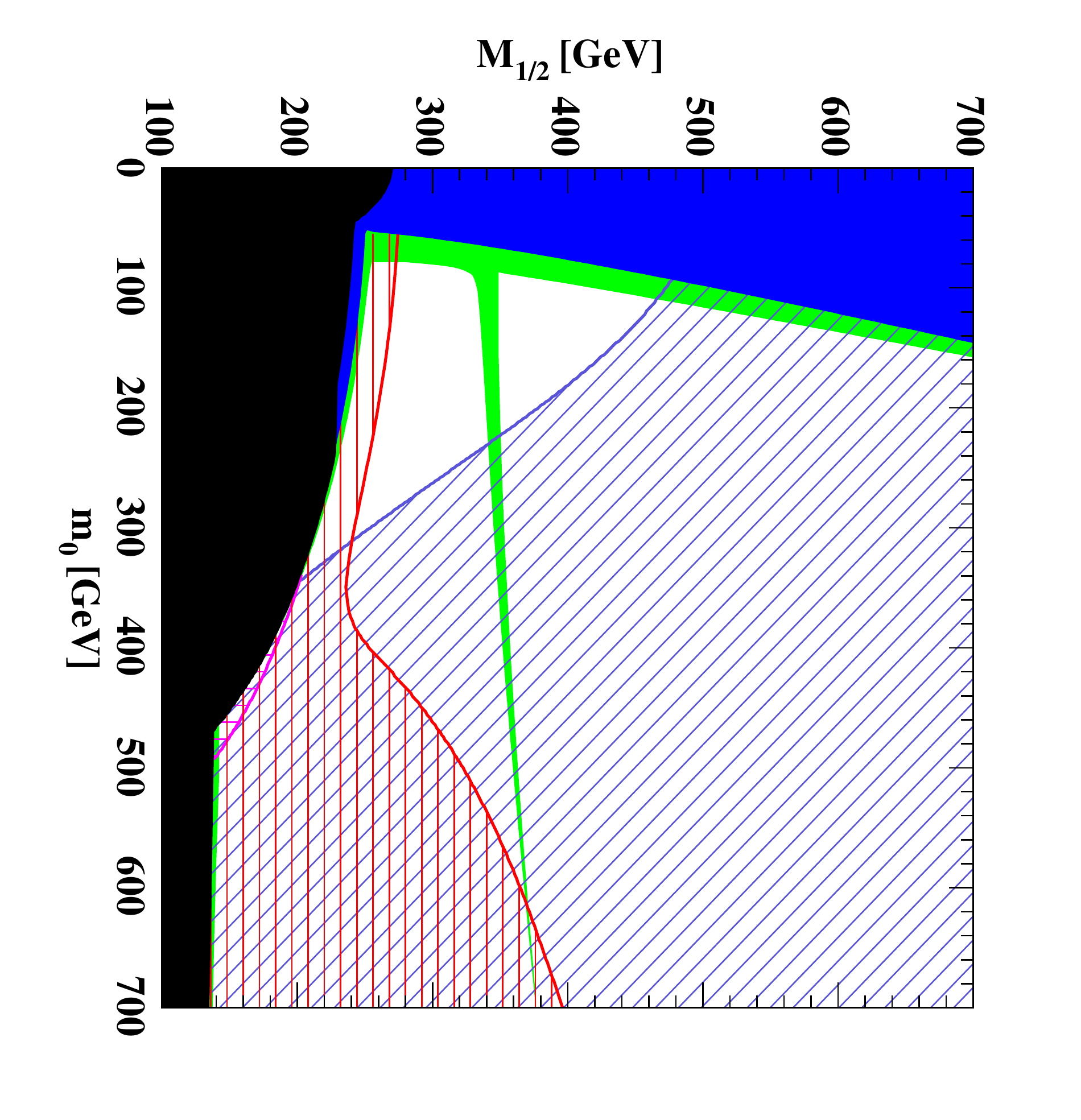}
\hspace*{-5mm}
\includegraphics[width=.5\textwidth, keepaspectratio,angle=90]{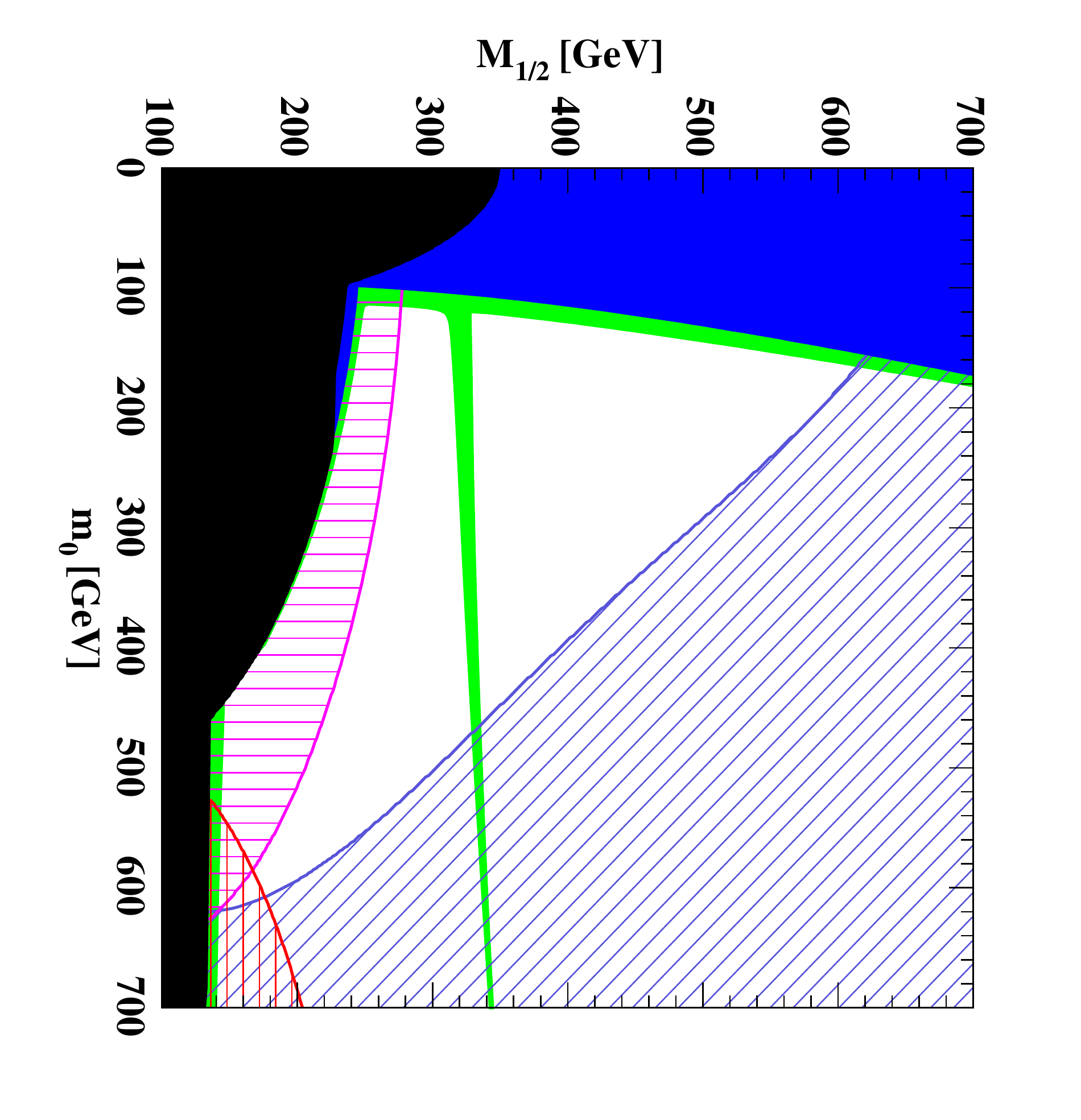}
\vspace*{-1.5cm}\\
\hspace*{7.3cm}(a)\hspace{7.7cm}(b)
\caption{\small Constraints in the $\m$, $\mhf$ plane for $\l=.1$, $\A=-900$~GeV, $\Ak=-60$~GeV and (a) $\tb=5$, (b) $\tb=10$. We show the region excluded by theoretical constraints or by LEP/Tevatron searches on sparticles (black), the region where a charged sfermion is the LSP (blue), the LEP limits from Higgs searches (red), the constraints from B-physics (pink) and from $\gmu$ (violet) . The region allowed by the upper bound on the DM relic density $\Omega h^2$ is displayed in green.}
\vspace*{-2mm}
\label{fig:tb5_10}
\end{figure}
%%%%%%%%%%%%%%%%%%%%%%%%%%%%%%%%%%%%%

As an example of a scenario with bino LSP annihilation through the $a$ resonance we consider in fig.~\ref{fig:tb5_10} the case $\A=-900$~GeV, $\Ak=-60$~GeV with (a) $\tb =5$, (b) $\tb =10$. The $a$ resonance is found around $\mhf\sim 350$~GeV, {\it ie} a bino mass $m_\bino\sim 143$~GeV and corresponds to $|2m_\bino-m_a| \lsim 3$~GeV. For larger negative values of $\Ak$, $m_a$ increases, as can be seen from eq.~(\ref{eq:singlets}), which means that rapid annihilation would be possible for a heavier bino LSP, that is a larger $\mhf$. However, when $m_\bino$ is large it becomes increasingly difficult to rely exclusively on Higgs exchange to have efficient enough annihilation. Thus, for large negative values of $\Ak$, the rapid annihilation region disappears. In addition, compatibility with $\amu$ requires not too large values for $\m$, $\mhf$, constraining further the allowed range of $\Ak \gsim -100$~GeV. On the other hand, if $|\Ak|$ is too small ($\Ak \gsim -20$~GeV), the $a$ resonance band appears at small values of $\mhf$ which are strongly constrained by chargino and Higgs searches at LEP at small $\tb$, and by $\bsg$ as one increases $\tb$. While $\Ak$ determines the value of $\mhf$ at which the $a$ resonance appears, the value of $\A$ steps in the Higgs and sfermion masses: if $|\A|$ is too small ($\A \gsim -500$~GeV) the LEP Higgs bounds become too strong and the whole $\m$, $\mhf$ plane is excluded. Similarly, if $|\A|$ is too large ($\A \lsim -1.5$~TeV), the sfermions become too light, excluding the $\m$, $\mhf$ region where $\amu$ is within the $2\sigma$ limits. Rapid annihilation of the bino LSP through the light scalar doublet $h$ can also occur at low values of $\mhf \sim 130$~GeV, as in the CMSSM. The low $\mhf$ region is however strongly constrained as explained above. Finally, as in the CMSSM, we find a bino-stau coannihilation band for values of $\m$ just above the stau LSP forbidden region (at small $\m$) as well as a narrow bino-stop coannihilation region at small values of $\m$ and $\mhf$ just above the stop LSP forbidden region (at $\mhf \sim 250$~GeV). The upper limit on $\ssi$ from CDMS~\cite{Ahmed:2008eu} is satisfied over the whole $\m$, $\mhf$ plane.

\begin{figure}[t!]
\vspace*{-5mm}
\hspace*{-5mm}
\includegraphics[width=.5\textwidth,keepaspectratio,angle=90]{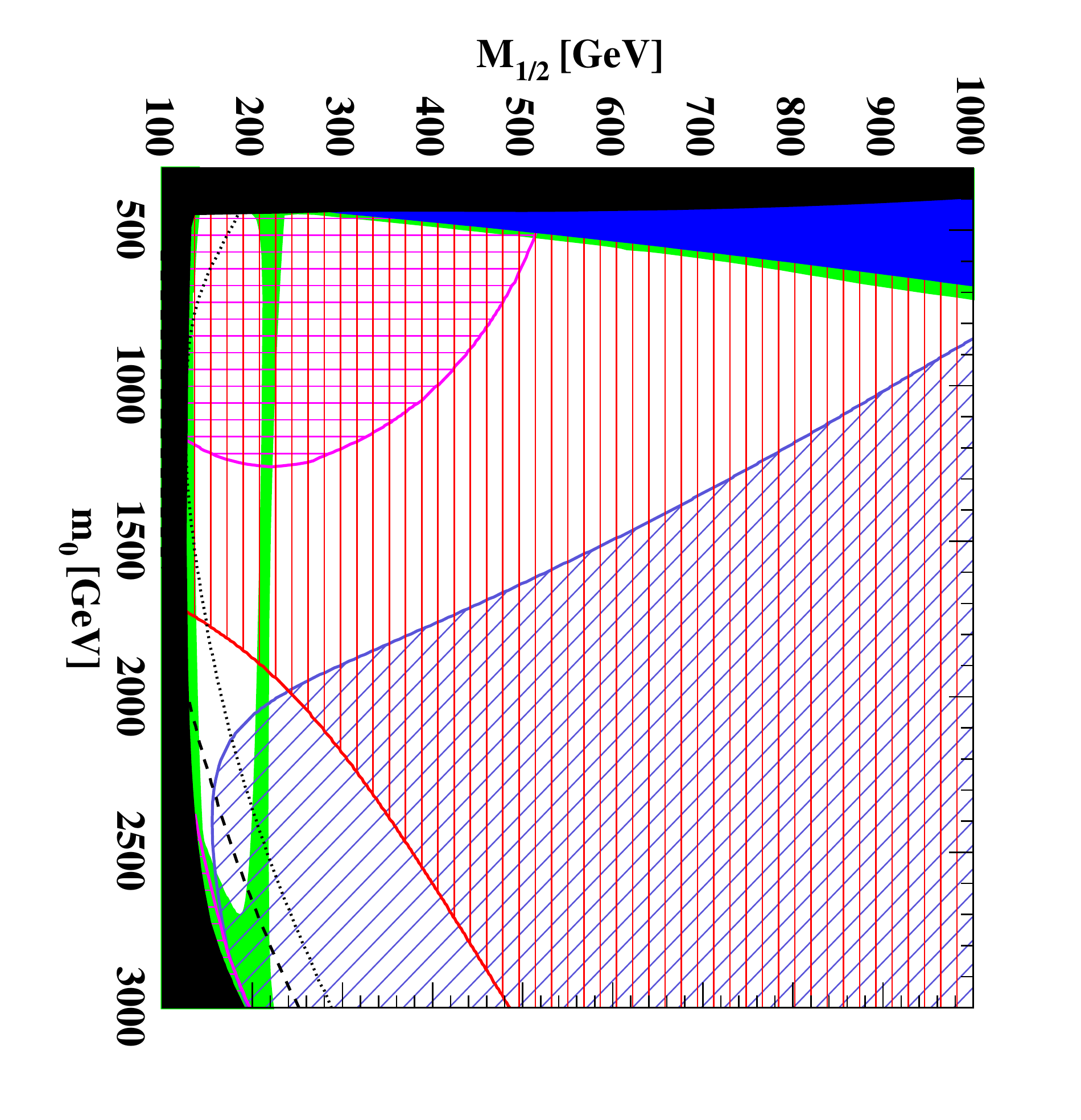}
\hspace*{-5mm}
\includegraphics[width=.5\textwidth, keepaspectratio,angle=90]{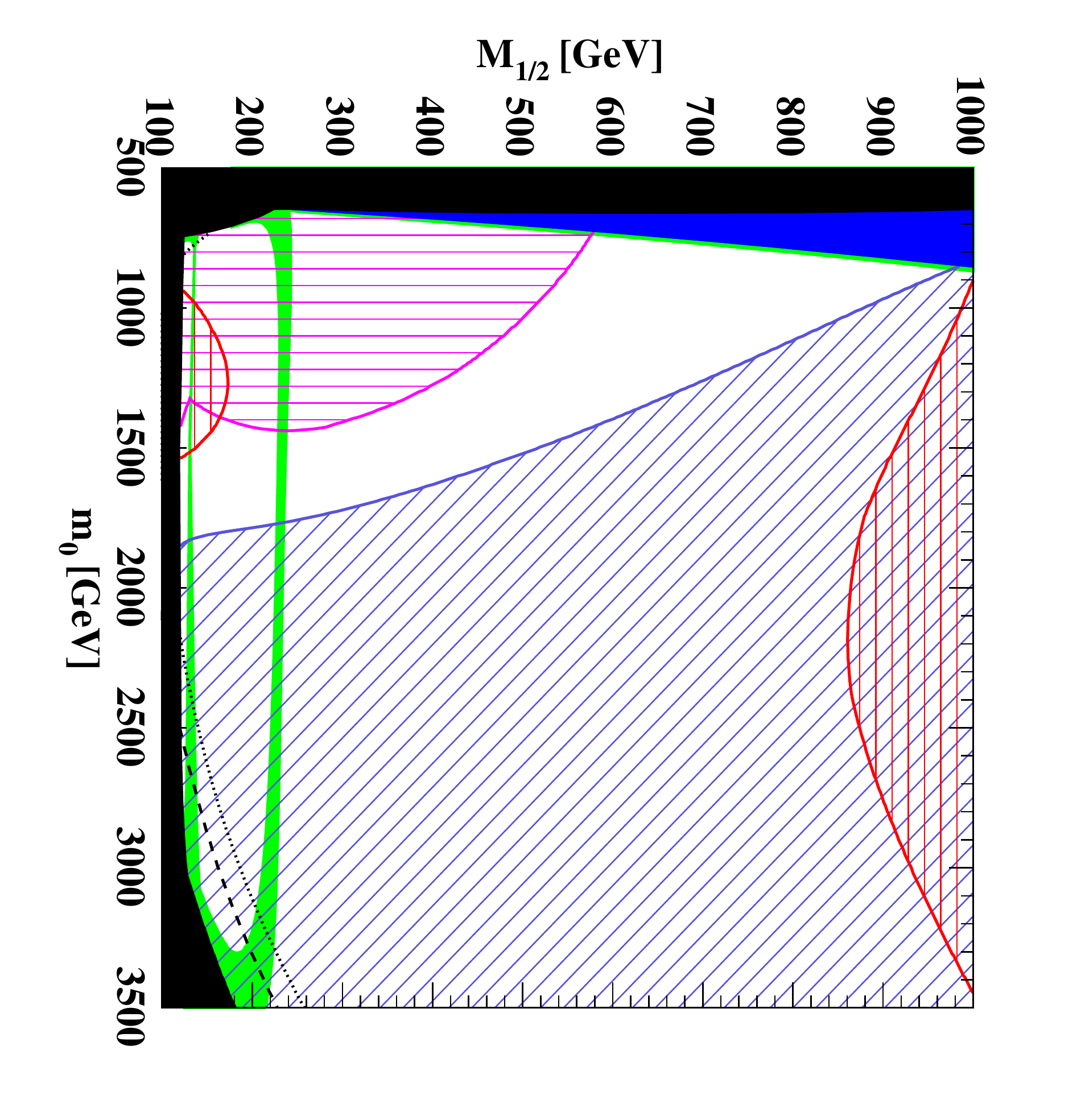}
\vspace*{-1.5cm}\\
\hspace*{7.3cm}(a)\hspace{7.7cm}(b)
\caption{\small Constraints in the $\m, \mhf$ plane for $\l=.1$, $\tb=50$, $\Ak=-60$~GeV and (a) $\A=-900$~GeV, (b) $\A=-1.5$~TeV. Same color code as in fig.~\ref{fig:tb5_10} with in addition the limit from CDMS on $\ssi$ for $(\sigma_{\pi N}, \sigma_0) = (55, 35)$~MeV (black dots) and $(45, 40)$~MeV (black dashes).}
\vspace*{-2mm}
\label{fig:tb50-}
\end{figure}

At large values of $\tb=50$, when $\A$, $\Ak < 0$, DM annihilation mechanisms are bino LSP annihilation through Higgs exchange, bino-sfermion coannihilation (at small $\m$) and higgsino LSP annihilation in the focus point region (large $\m$) . This is illustrated in fig.~\ref{fig:tb50-} for $\Ak = -60$~GeV and (a) $\A = -900$~GeV, (b) $\A = -1.5$~TeV. The possible Higgs resonances are again the light scalar doublet $h$ at low $\mhf \sim 130$~GeV, just above the chargino exclusion limit from LEP or the pseudoscalar singlet $a$ for slightly larger values of $\mhf \sim 200$~GeV. The B-physics observables (predominantly $\bsg$) restricts the small $\m$, $\mhf$ region, whereas the $\gmu$ constraint restricts the large $\m$, $\mhf$ region eliminating most the focus point region. What is left of the focus point region by the $\gmu$ constraint at large $\m$ is excluded by B-physics constraints (here, predominantly $\bsmu$ and $\btau$). The LEP constraints on the Higgs sector excludes a large region of the parameter space, up to $\m \sim 1.7$~TeV for $\A = -900$~GeV, eliminating all the bino-stau coannihilation region and leaving only a small region of resonance annihilation. However, over the whole region excluded by Higgs searches in fig.~\ref{fig:tb50-}(a), the lightest Higgs doublet mass is just below the LEP limit ($114.5$~GeV). Hence, this constraint is easily relaxed by taking a larger value of $|\A|$, ($\A=-1.5$~TeV in fig.~\ref{fig:tb50-}(b)) or by assuming a $3$~GeV theoretical uncertainty on the Higgs mass. Increasing $|\A|$ also somewhat strengthtens the $\gmu$ constraint by increasing the mixing in the smuon sector. The influence of $\A$ is most noticeable in the focus point region when $\mu$ is not too large since the smuon mixing is proportionnal to $A_\mu-\mu\tb$. In the region where the heavy Higgs doublet $H$ is light, the predictions for DM direct detection rate can be large. The CDMS limit thus constrains small values of $\mhf$. Some area of the parameter space is excluded even when being conservative and fixing $(\sigma_{\pi N}, \sigma_0)=(45,40)$~MeV, a choice that gives a small coefficient for the $s$ quark content in the nucleon.

%%%%%%%%%%%%%%%%%%%%%%%%%%%%%%%%%%%%%
%%%%%%%%%%%%%%%%%%%%%%%%%%%%%%%%%%%%%

\begin{figure}[t!]
\vspace*{-5mm}
\hspace*{-5mm}
\includegraphics[width=.5\textwidth,keepaspectratio,angle=90]{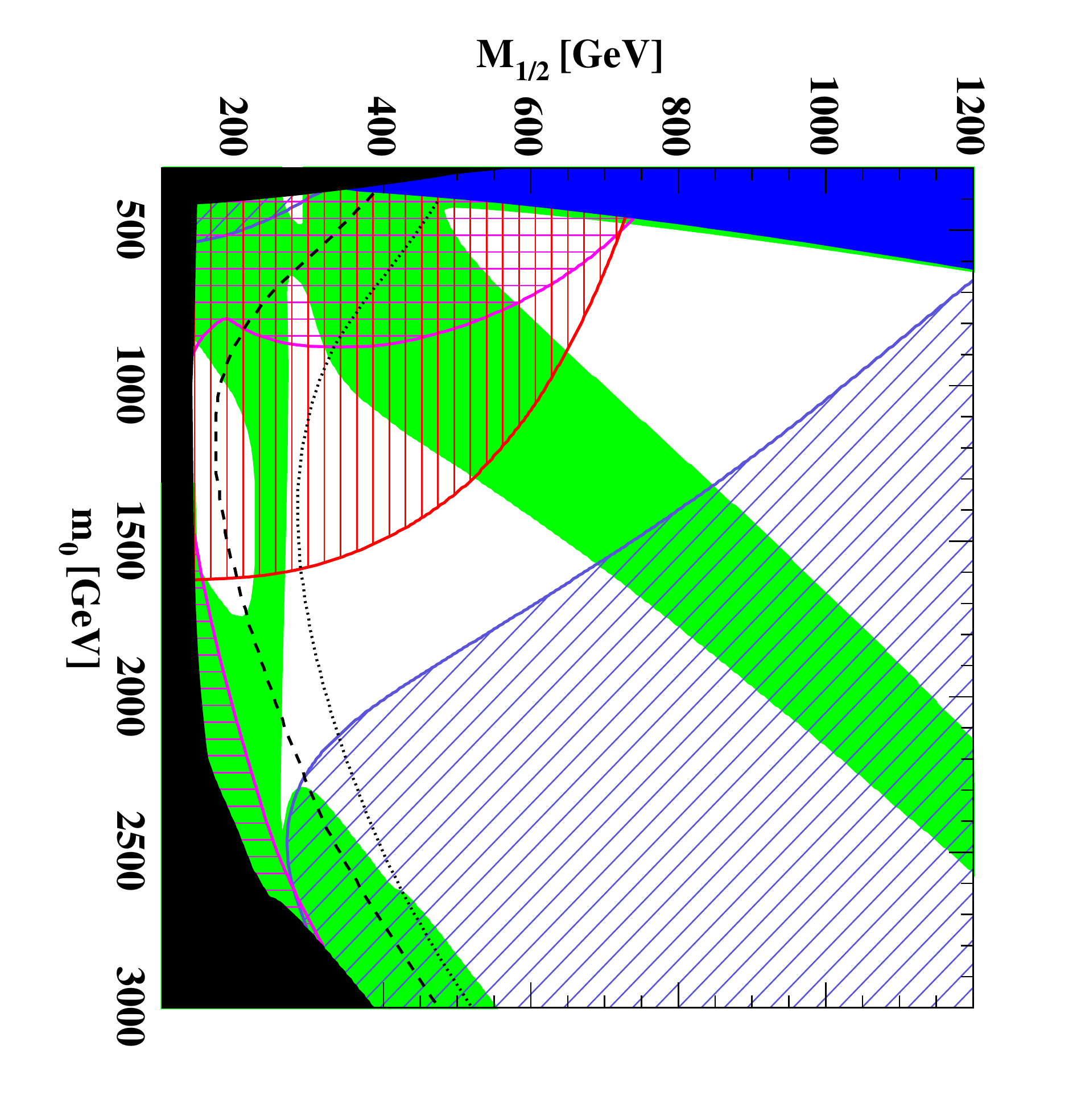}
\hspace*{-5mm}
\includegraphics[width=.5\textwidth, keepaspectratio,angle=90]{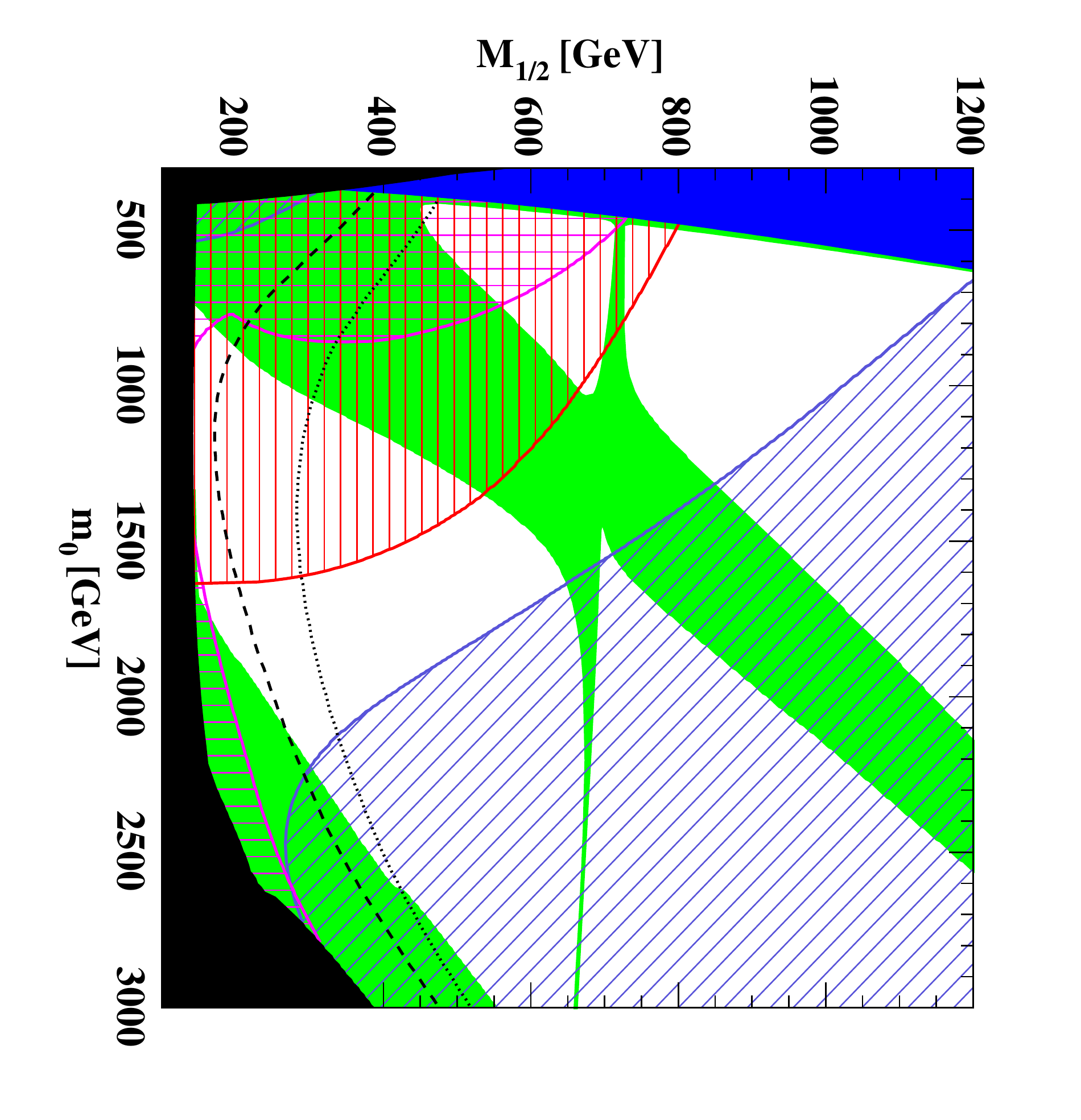}
\vspace*{-1.5cm}\\
\hspace*{7.3cm}(a)\hspace{7.7cm}(b)
\caption{\small Constraints in the $\m, \mhf$ plane for $\l=.1$, $\tb=50$, $\A=900$~GeV and (a) $\Ak=60$~GeV, (b) $\Ak=240$~GeV. Same color code as in fig.~\ref{fig:tb5_10} and fig.~\ref{fig:tb50-}.}
\vspace*{-2mm}
\label{fig:tb50+}
\end{figure}

For large $\tb = 50$ and $\A$, $\Ak >0$, all Higgs states, except the scalar singlet $s$, can contribute significantly to the bino LSP annihilation. In fig.~\ref{fig:tb50+}, we take $\A=900$~GeV and (a) $\Ak = 60$~GeV, (b) $\Ak = 240$~GeV. The possible Higgs resonances are: (i)~the light scalar doublet $h$ just above the chargino exclusion limit from LEP at $\mhf \sim 130$~GeV; (ii)~the pseudoscalar singlet $a$ for $\mhf \sim 200$~GeV when $\Ak = 60$~GeV and $\mhf \sim 700$~GeV when $\Ak = 240$~GeV; (iii)~the heavy scalar/pseudoscalar doublets $H/A$ in the wide diagonal band. In addition, for such large values of $\tb$, as in the CMSSM, one finds a region at large $\m$ where the LSP has a significant higgsino component and therefore annihilates efficiently. Finally, at small $\m$, one still has a thin band where the bino LSP coannihilates with the stau LSP along the excluded region where the stau is the LSP. Note that regions at small $\m$, $\mhf$ are in conflict with the LEP Higgs constraints as well as with the $\bsmu$ and/or $\btau$ constraints (recall that for $\A>0$ the $\bsg$ constraint is less effective). Regions at large $\m$ and small $\mhf$ are also excluded by the $\bsmu$ and/or $\btau$ constraints since the heavy neutral/charged Higgs doublets are rather light. Note that as one increases $\m$ for a fixed value of $\mhf$, say $200$~GeV, the charged Higgs mass $m_{H^\pm} \sim m_A$ decreases until $Br(\btau)$ drops below the allowed range for $\m \sim 2$~TeV because of the destructive interference of the $H^\pm$ contribution with the standard model $W^\pm$ contribution. When $\m$ is increased further $m_{H^\pm}$ continues to decrease such that $Br(\btau)$ is large and dominated by the $H^\pm$ contribution. Hence, there is a narrow region where this constraint is satisfied, although for that region the branching ratio for $Br(\bsmu)$ is too large (in fig.~\ref{fig:tb50+} only the combined exclusion from B-physics observables is displayed). Then, as one increases $\m$, $m_{H^\pm} \sim m_A$ decreases until it falls below the LEP limit for $\m \gsim 2.4$~TeV and eventually becomes even negative for larger values of $\m$. The constraint on $\gmu$ excludes two distinct regions: at small $\m$, $\mhf$ the prediction for $\amu$ is too large, and at large $\m$, $\mhf$ it is below the preferred range. The direct detection rate exceeds the limit from CDMS both when the LSP has a large higgsino component (the large $\m$ focus point region) and/or when the heavy Higgs doublet $H$ is light (at small $\mhf$).

%%%%%%%%%%%%%%%%%%%%%%%%%%%%%%%%%%%%%
%%%%%%%%%%%%%%%%%%%%%%%%%%%%%%%%%%%%%

Setting the top quark mass at the minimum of the $2\sigma$ range, $m_t=171.2$~GeV, one finds that a much larger area of the $\m$, $\mhf$ plane satisfies the cosmological upper bound on the relic density when $\A, \Ak > 0$. This is because decreasing $m_t$, all other parameters fixed, one obtains both a lower value for $\mu$, hence a larger higgsino fraction for the LSP, as well as a lighter heavy Higgs doublet sector. The focus point and the $H/A$ funnel regions of fig.~\ref{fig:tb50+} then merge in a large allowed band. This effect was already observed in~\cite{Hugonie:2007vd}, fig.~3(b) for a different choice of parameters. Furthermore, since $\mu$ is smaller, the smuon mixing is also smaller and the $\gmu$ constraint is less severe, especially in the focus point region. Even though the B-physics and LEP Higgs constraints are slightly more severe and a LSP with a larger higgsino fraction necessarily implies a larger direct detection rate, often exceeding the CDMS limit, one finds a larger area of the $\m$, $\mhf$ plane that satisfies all constraints. Conversely, increasing $m_t=174$~GeV implies that the focus point region is pushed at higher values of $\m$ and the $H/A$ funnel appears at higher values of $\mhf$, separating further the two DM allowed bands of fig.~\ref{fig:tb50+} . Similarly, B-physics and LEP Higgs constraints are slightly less severe when $m_t$ is larger, while $\gmu$ constraints are strengthened, moving the allowed area between these constraints in fig.~\ref{fig:tb50+} to smaller values of $\m$, $\mhf$. The only remaining regions passing all constraints in this case are the bino-stau coannihilation region as well as the three Higgs funnels -- the heavy doublet $H/A$, the light doublet $h$ and the pseudoscalar singlet $a$, the latter two being however disfavoured by the CDMS constraints. Varying $m_t$ when $\A, \Ak < 0$ has a similar effect on the location of the focus point region and on the phenomenological constraints. Yet in this case, there is no $H/A$ funnel and the region in the $\m$, $\mhf$ plane that satisfies all constraints is only shifted, but not much larger for $m_t=171.2$~GeV.

%%%%%%%%%%%%%%%%%%%%%%%%%%%%%%%%%%%%%
\subsection{\bf Small $\l$: singlino LSP}\label{sec:sing}
%%%%%%%%%%%%%%%%%%%%%%%%%%%%%%%%%%%%%

The singlino LSP can be found only for $\l \ll 1$, this means $\l=.01$ in our selected scans (qualitatively, the results are similar for smaller values of $\l$). In this case, we have an effective MSSM with an almost decoupled singlet sector and we do not expect to have singlet Higgs resonances. Thus the only mechanism that can provide the correct relic density for a singlino LSP is coannihilation, or more precisely self-annihilation of the NLSP. This works most efficiently when the LSP and NLSP are well below the TeV scale. We therefore expect to find WMAP allowed regions for choices of input parameters that predict a singlino LSP at small values of $\m$ where the stau is light, or at large $\m$ and $\tb$, in the focus point region, where the higgsino is light. In the small singlet mixing limit ($\l \ll 1$) the singlino mass is given by $m_\singlino = 2 \nu$ which, according to eq.~(\ref{eq:def}), implies
\beq\label{eq:sing}
m_\singlino = 2(B-\Al) \, .
\eeq
Hence, the singlino mass depends mainly on $\m$, $\mhf$ (through $B$, see eq.~(\ref{eq:min})) and $\A$ (through $\Al$ computed by integration of the RGEs) but is mostly independent of $\Ak$. It then suffices to take $\Ak = \pm 50$ GeV (depending on the sign of $\A$) in order to have positive mass squared for the Higgs singlet states.

%%%%%%%%%%%%%%%%%%%%%%%%%%%%%%%%%%%%%
\begin{figure}[t!]
\vspace*{-5mm}
\hspace*{-5mm}
\includegraphics[width=.5\textwidth,keepaspectratio,angle=90]{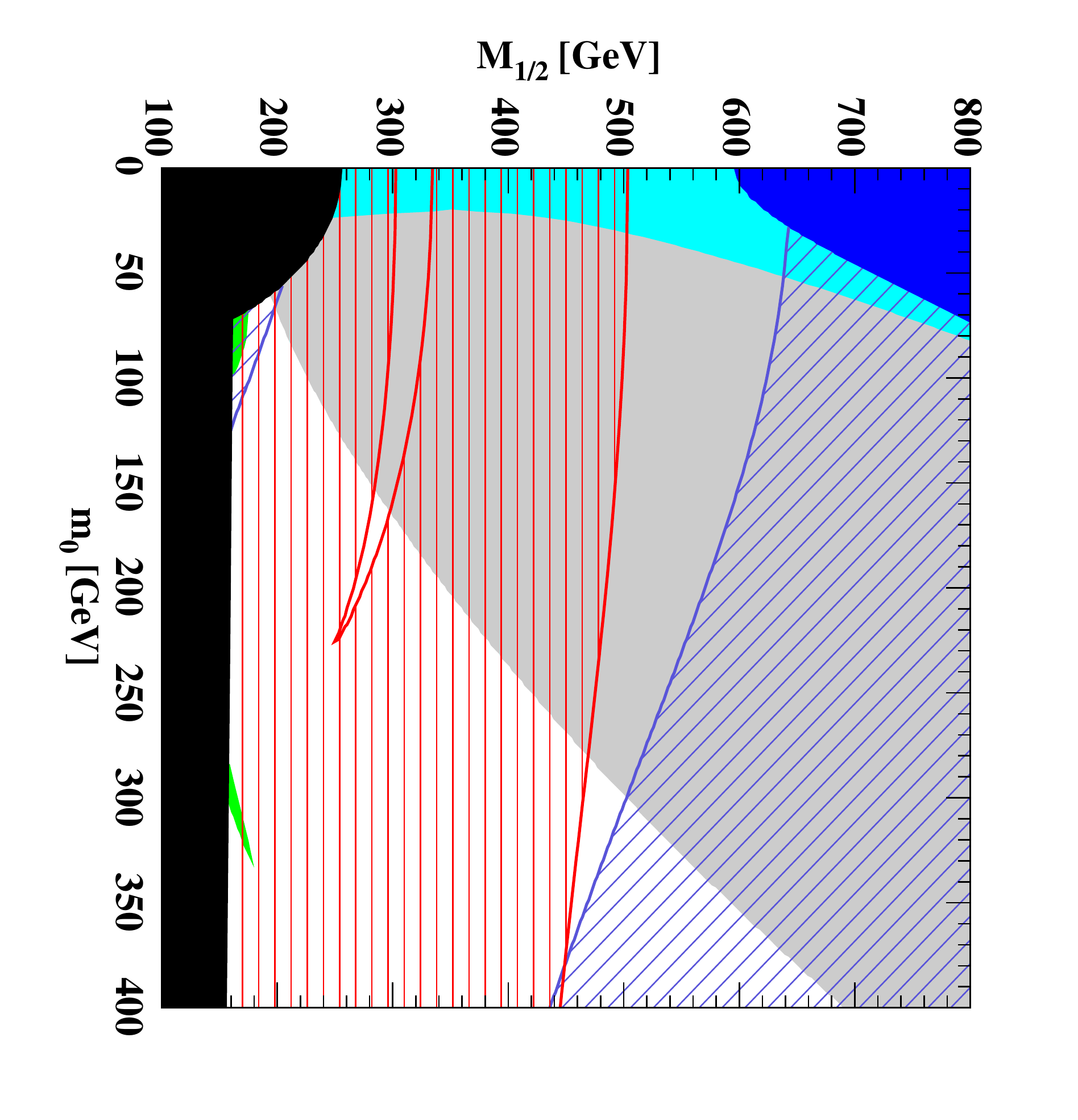}
\hspace*{-5mm}
\includegraphics[width=.5\textwidth, keepaspectratio,angle=90]{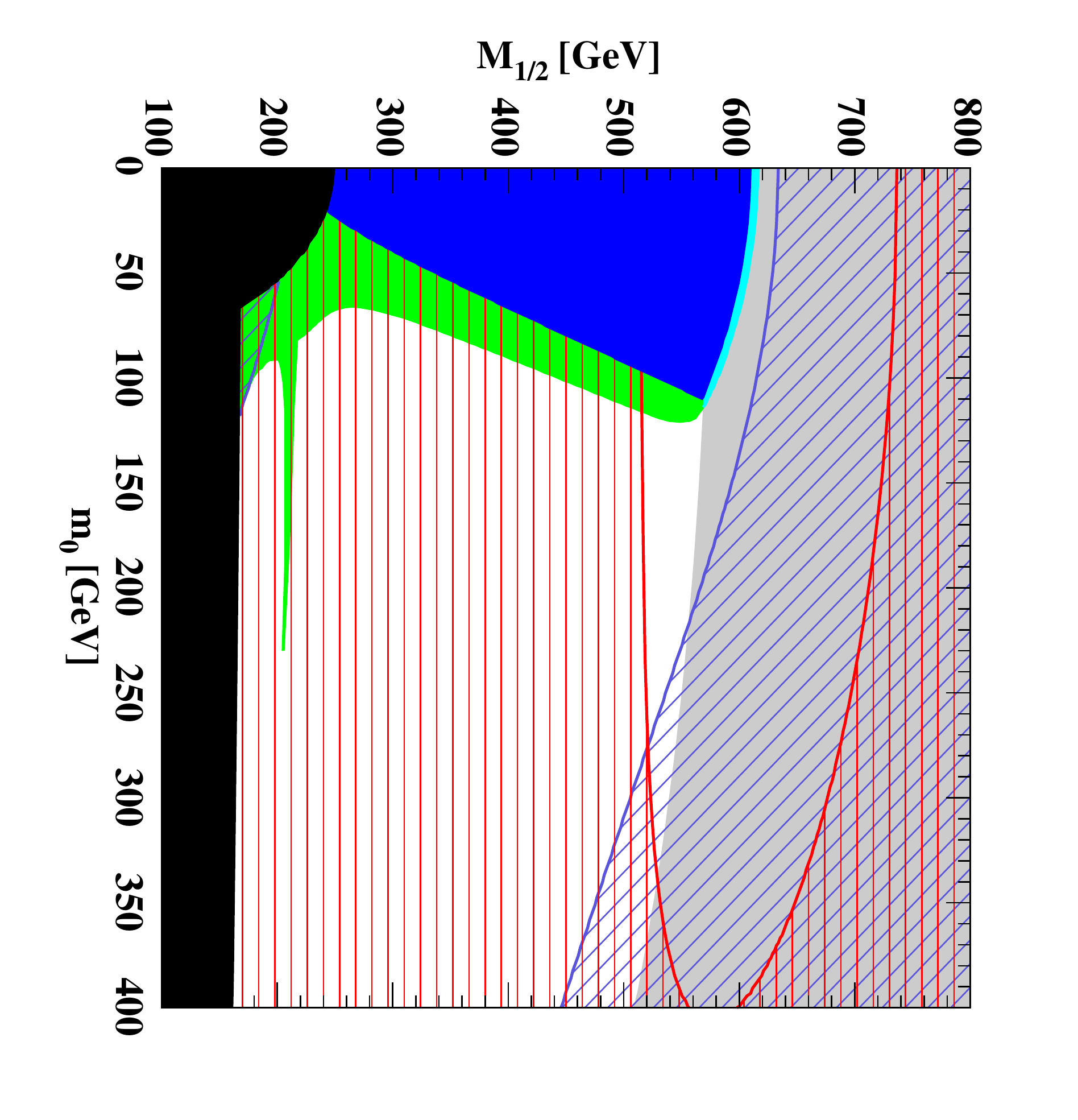}
\vspace*{-1.5cm}\\
\hspace*{7.3cm}(a)\hspace{7.7cm}(b)
\caption{\small Constraints in the $\m$, $\mhf$ plane for $\l = .01$, $\tb = 10$ and (a) $\A = -30$~GeV, $\Ak = -50$~GeV, (b) $\A = 300$~GeV, $\Ak = 50$~GeV. Same color code as fig.~\ref{fig:tb5_10}, with in addition: the singlino LSP region with relic density above the WMAP limit (grey) and below WMAP limit (cyan).}
\vspace*{-2mm}
\label{fig:singlino_10}
\end{figure}
%%%%%%%%%%%%%%%%%%%%%%%%%%%%%%%%%%%%%

For intermediate values of $\tb = 5$ or $10$, there is no focus point region and we can restrict ourselves to small values of $\m$ where the singlino-stau coannihilation is expected to work out. In this region of parameter space however, it is difficult to satisfy both the LEP Higgs constraints, which exclude small values of $\mhf$, and the $\gmu$ constraint, which exclude large values of $\mhf$. For $\tb = 5$, we have checked, by fixing $\m = 0$ and scanning over the only two remaining parameters, $\mhf$ and $\A$, that the excluded regions overlap, leaving no possibility. We have repeated the same exercise for $\tb = 10$ and found that for $\A = -30$~GeV, $\Ak = -50$~GeV and $\A = 300$~GeV, $\Ak = 50$~GeV there are singlino LSP regions with the correct amount of DM, satisfying both the LEP Higgs and the $\gmu$ constraints, see fig.~\ref{fig:singlino_10}. Note that the precise value of $\A$ is crucial for these plots.

For $\Ak < 0$ (fig.~\ref{fig:singlino_10}(a)), the singlino mass, eq.~(\ref{eq:sing}), is positive (see eq.~(\ref{eq:rel})) and grows with $\m$ and $\mhf$. Therefore, one finds a singlino LSP region at small $\mhf$, below the stau LSP forbidden region. At the frontier of the two regions, one finds a band where the singlino LSP coannihilates with the stau NLSP, giving the correct amount of DM. For smaller values of $\mhf$, the singlino LSP coannihilates with the stop NLSP. Large values of $\mhf \gsim 650$~GeV are excluded by the $\gmu$ constraints, while the LEP Higgs constraints exclude small values of $\mhf \lsim 500$~GeV. In the small region that passes the latter constraints near $\mhf=300$~GeV, the light doublet $h$ and the singlet $s$ are almost degenerate in mass, and mixing effects push $m_h$ just above the LEP limit. The LEP Higgs constraints can however be relaxed if one chooses a smaller value of $\l$ (e.g. $\l = .001$) or a larger value of $m_t$, both solutions having the same effect of increasing $m_h$. Alternatively, one could allow for a $3$~GeV theoretical uncertainty on $m_h$, in which case the LEP Higgs excluded region reduces drastically.

For $\Ak > 0$ (fig.~\ref{fig:singlino_10}(b)), the singlino mass is negative and grows with $\m$ and $\mhf$. Therefore, the singlino LSP is found at large values of $\mhf$, below the excluded region where the pseudoscalar singlet mass squared becomes negative (occurring at $\mhf \gsim 1.2$~TeV). At the frontier with the forbidden stau LSP region, one finds a region where the relic density of the singlino LSP is below the WMAP limit. For our choice of $\A = 300$~GeV, this region satisfies in addition both the LEP Higgs and the $\gmu$ constraints. However, the same remarks concerning the LEP Higgs constraints as in the case $\Ak<0$ apply here. In all cases, the mass difference between the singlino and the stau must be $\lsim 3$~GeV for the coannihilation mechanism to be efficient enough.

%%%%%%%%%%%%%%%%%%%%%%%%%%%%%%%%%%%%%
\begin{figure}[t!]
\vspace*{-5mm}
\hspace*{-5mm}
\includegraphics[width=.5\textwidth,keepaspectratio,angle=90]{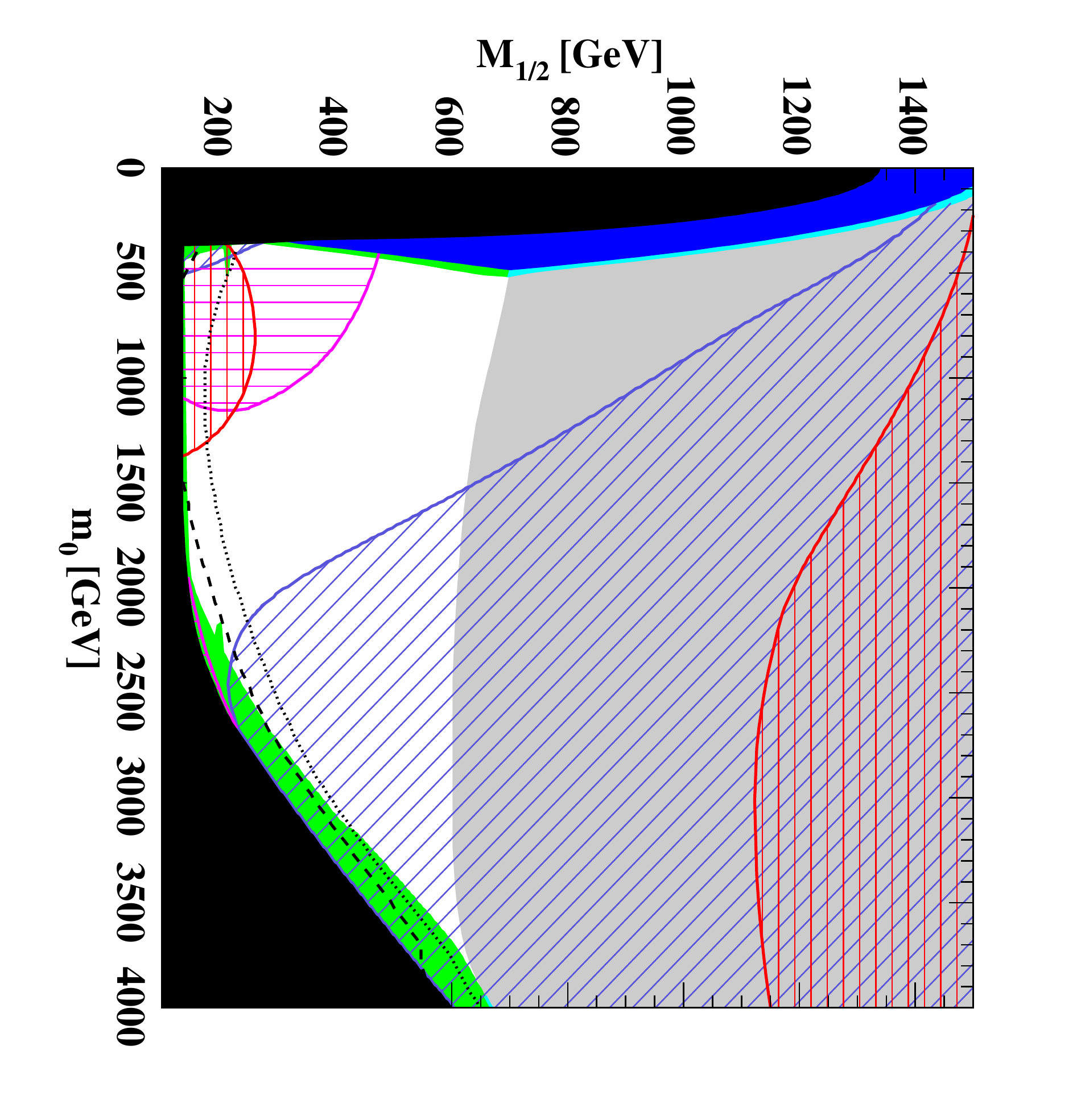}
\hspace*{-5mm}
\includegraphics[width=.5\textwidth, keepaspectratio,angle=90]{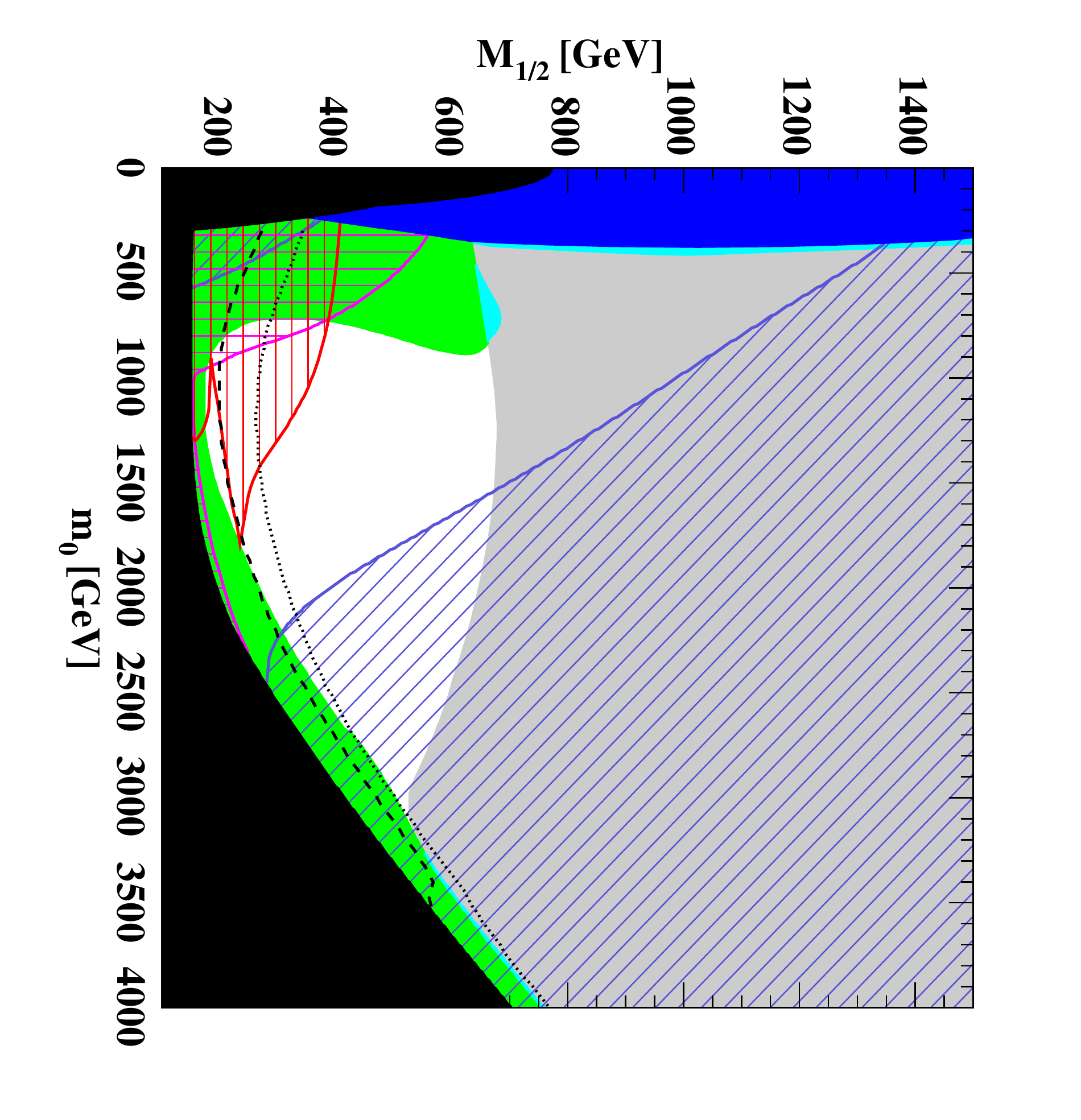}
\vspace*{-1.5cm}\\
\hspace*{7.3cm}(a)\hspace{7.7cm}(b)
\caption{\small Constraints in the $\m$, $\mhf$ plane for $\l = .01$, $\tb = 50$ and (a) $\A = -600$~GeV, $\Ak = -50$~GeV, (b) $\A = 0$~GeV, $\Ak = 50$~GeV. Same color code as fig.~\ref{fig:singlino_10}.}
\vspace*{-2mm}
\label{fig:singlino_50}
\end{figure}
%%%%%%%%%%%%%%%%%%%%%%%%%%%%%%%%%%%%%

For large $\tb = 50$, $B \simeq 0$ (see eq.(\ref{eq:min})) and the singlino mass is $m_\singlino \simeq -2\Al$ depending mainly on $\A$. Hence, for large values of $\mhf$ the bino is heavy and the singlino is the LSP. In addition, both the LEP Higgs and the $\gmu$ constraints are much weaker for large values of $\tb$. In fig.~\ref{fig:singlino_50} we present the results for $\A=-600$~GeV, $\Ak=-50$~GeV and $\A = 0$, $\Ak = 50$~GeV. In both cases, the singlino is the LSP for $\mhf \gsim 700$~GeV and its relic density is below the WMAP upper bound when it coannihilates with the stau NLSP at low $\m$. For $\mhf \lsim 700$~GeV, as in the CMSSM the bino LSP also coannihilates with the stau NLSP at low $\m$ or annihilates through the light Higgs doublet $h$ resonance when $\mhf\sim130$~GeV. The latter region is however mostly excluded by the B-physics and LEP Higgs constraints. As in the CMSSM, we find a WMAP allowed band at large $\m$, where the LSP has a significant higgsino component and therefore annihilates efficiently. Most of this band is however excluded by CDMS, even when taking the conservative limit. At the frontier between this region and the singlino LSP region, we find a thin WMAP allowed band where the singlino LSP coannihilates with the higgsino. Yet, this region is incompatible with the $\gmu$ constraint. For $\Ak > 0$, the heavy Higgs doublet $H/A$ resonances can also contribute to the annihilation of the bino LSP in a large diagonal band, as in fig.~\ref{fig:tb50+}. However, in the singlino LSP region, when $\mhf \gsim 700$~GeV, this mechanism is not sufficient to bring the relic density below the WMAP upper bound, except in a thin band at the frontier between bino and singlino LSP regions where the singlino LSP coannihilates with the bino NLSP which in turn annihilates through the $H/A$ resonance.
\pagebreak

%=======================================================================
\section{Prospects for dark matter direct detection}\label{sec:dd}
%=======================================================================

In sec.~\ref{sec:res} we have seen that direct detection experiments (CDMS) already constrain the parameter space of the CNMSSM at large values of $\tb$, especially in the focus point region at large $\m$. Here we present in more details the predictions for the SI cross sections for elastic scattering of neutralinos on protons in scenarios with an extra Higgs resonance. Cross sections for direct detection of a singlino LSP are always far too small to be within the present or future experimental reach. Recall that experimental sensitivities $\ssi\approx 10^{-9}-10^{-8}$~pb are expected within a few years and that the planned ton-scale detectors such as Warp, Xenon, Eureca~\cite{Kraus:2006pj} or SuperCDMS~\cite{Akerib:2006rr} will reach a sensitivity around $10^{-10}$~pb within a decade. We will see that for large regions of the parameter space of the CNMSSM the predictions for SI direct detection cross sections are well within reach of the ton-scale detectors. In many cases signals are to be expected much before that. We take here the \micro\ default values, $(\sigma_{\pi N}, \sigma_0) = (55, 35)$~MeV to determine the quark coefficients in the nucleon. Almost one order of magnitude variation in the prediction can result from diffrent choices for the quark coeficients~\cite{Bottino:2001dj,Belanger:2008sj,Ellis:2008hf}.

\begin{figure}[t!]
\vspace*{-5mm}
\hspace*{-5mm}
\includegraphics[width=.5\textwidth,keepaspectratio,angle=90]{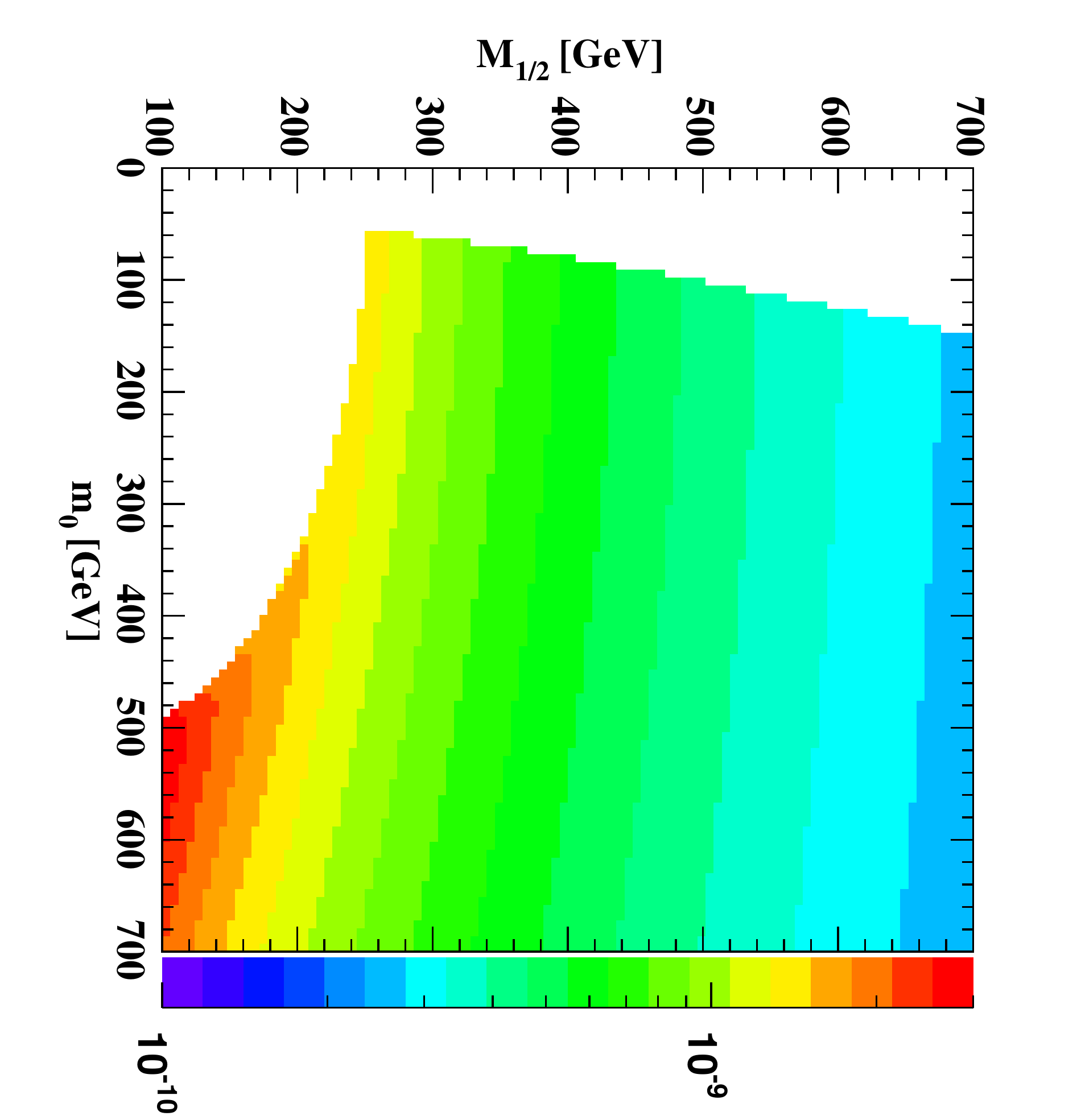}
\hspace*{-5mm}
\includegraphics[width=.5\textwidth, keepaspectratio,angle=90]{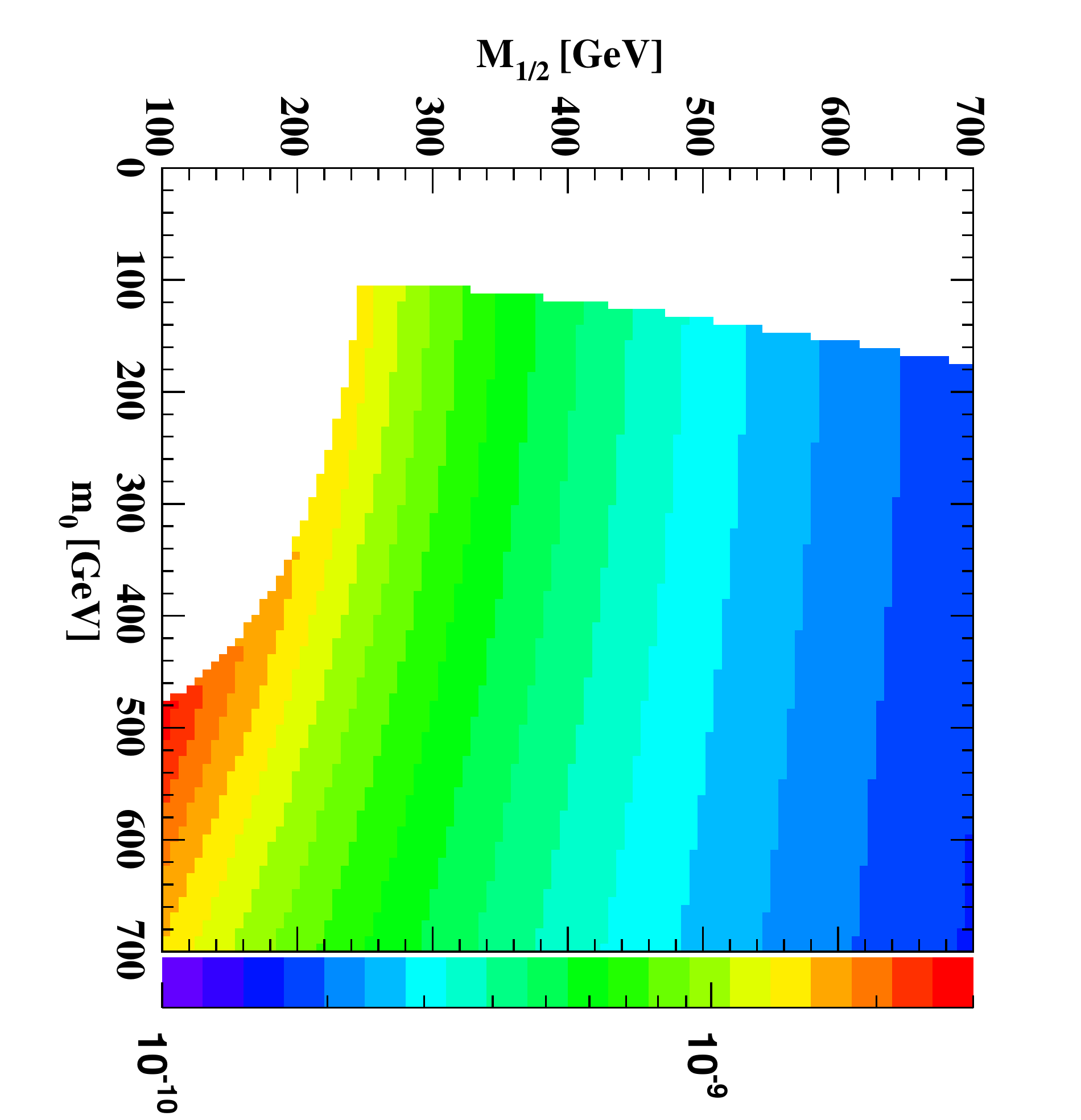}
\vspace*{-1.5cm}\\
\hspace*{7.3cm}(a)\hspace{7.7cm}(b)
\caption{\small Contours of $\ssi$ (in pb) in the $\m$, $\mhf$ plane for $\l=.1,\A=-900$~GeV, $\Ak=-60$~GeV and (a) $\tb=5$, (b) $\tb=10$.}
\vspace*{-2mm}
\label{fig:direct_tb510}
\end{figure}

At small to intermediate values of $\tb$, $\ssi$ is dominated by the light Higgs doublet $h$ exchange although both the heavy Higgs doublet $H$ and squarks exchange diagrams also contribute. Contour plots of $\ssi$ for the cases considered in the previous section, $\l=.1$, $\A=-900$~GeV, $\Ak=-60$~GeV with $\tb=5$ and $10$, are displayed in fig.~\ref{fig:direct_tb510}. (In the regions where no contours are displayed, either the stop or the stau is the LSP). In the region favoured by all constraints (see fig.~\ref{fig:tb5_10}), the SI cross section is predicted to be between $5(2)$ and $10\times 10^{-10}$~pb for $\tb=5(10)$ with a LSP mass between $70$ and $200(250)$~GeV (here the bino LSP mass is simply given by $m_\bino \simeq .41 \mhf$), within reach of the ton-scale detectors. Note that typically $\ssi$ decreases with $\mhf$, as both the mass of the heavy Higgs doublet and of the squarks increase with $\mhf$. In the bino-stau coannihilation region at small $\m$, large $\mhf$ (which is however disfavoured by the $\gmu$ constraint) the SI cross section drops below the sensitivity of ton-scale detectors as in the CMSSM.

%%%%%%%%%%%%%%%%%%%%%%%%%%%%%%%%%%%%
%%%%%%%%%%%%%%%%%%%%%%%%%%%%%%%%%%%%%

\begin{figure}[t!]
\vspace*{-5mm}
\hspace*{-5mm}
\includegraphics[width=.5\textwidth,keepaspectratio,angle=90]{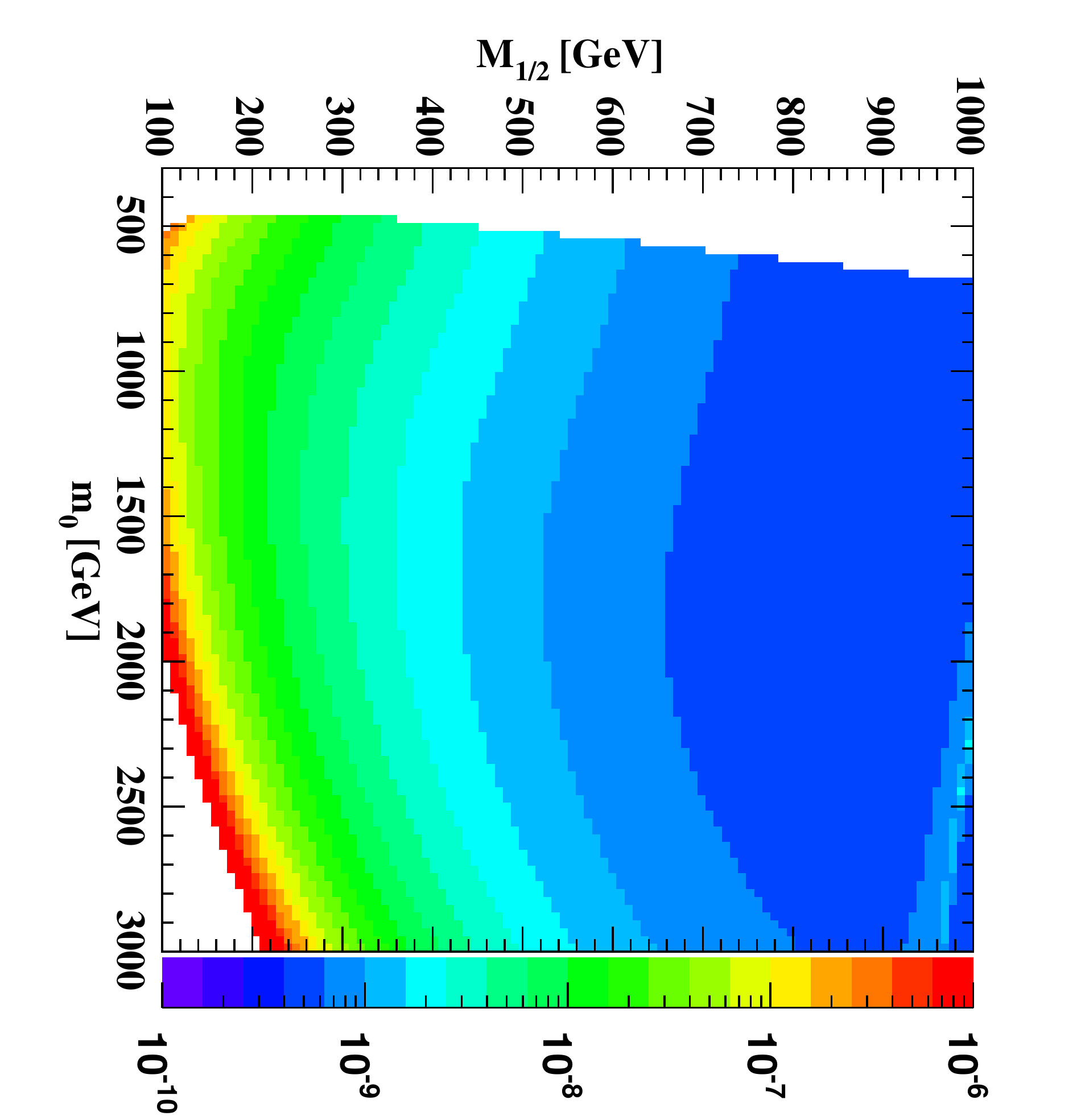}
\hspace*{-5mm}
\includegraphics[width=.5\textwidth, keepaspectratio,angle=90]{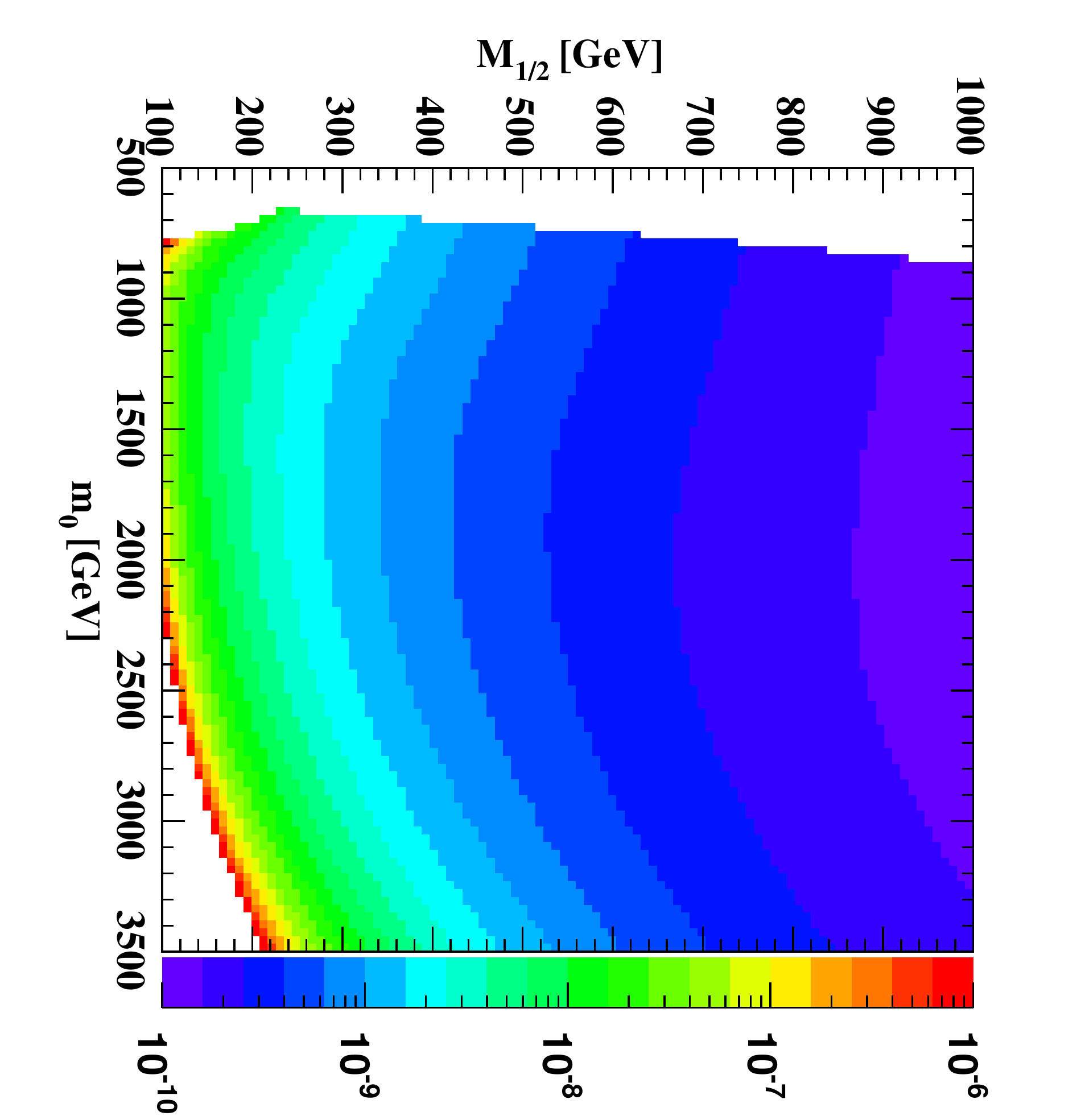}
\vspace*{-1.5cm}\\
\hspace*{7.3cm}(a)\hspace{7.7cm}(b)
\caption{\small Contours of $\ssi$ (in pb) in the $\m$, $\mhf$ plane for $\l=.1$, $\tb=50$, $\Ak=-60$~GeV and (a) $\A=-900$~GeV, (b) $\A=-1.5$~TeV.}
\vspace*{-2mm}
\label{fig:direct_tb50-}
\end{figure}

For $\tb=50$, $\ssi$ is dominated by the heavy Higgs doublet $H$ exchange since the couplings of $H$ to quarks and to the LSP are $\tb$ enhanced. This enhancement compensates for the mass suppression factor $m_{h}^2/m_{H}^2$. Cross sections near or above the present limit are obtained for small $\mhf$, see fig.~\ref{fig:direct_tb50-}(a) for $\l=.1$, $\A=-900$~GeV and $\Ak=-60$~GeV. In the allowed region where the LSP annihilates through the pseudoscalar singlet $a$ resonance ($\mhf\sim 200$~GeV, see fig.~\ref{fig:tb50-}) we find $\ssi\approx 2\times 10^{-8}$~pb. This region should soon be probed experimentally. A large negative value for $\A=-1.5$~TeV induces larger heavy Higgs masses, thus a smaller SI cross section, see fig.~\ref{fig:direct_tb50-}(b). In this case, the pseudoscalar singlet funnel corresponds to $\ssi\approx 2.5\times 10^{-9}$~pb. In both plots of fig.~\ref{fig:direct_tb50-}, regions where no contours are displayed correspond either to regions where no solution to minimization eqs.~(\ref{eq:min}) is found (large $\m$), the LSP is a sfermion (small $\m$). In the bino-stau coannihilation allowed region (small $\m$, $600$~GeV$\lsim\mhf\lsim 1$~TeV, see fig.~\ref{fig:tb50-}) the SI cross section is found to be barely within the reach of the ton-scale detectors in both cases. On the contrary, as mentioned in sec.~\ref{sec:res}, the predictions for $\ssi$ are already above the present experimental bounds for most of the focus point region (large $\m$, small $\mhf$) unless one allows for reduced coefficient of the $s$ quarks in the nucleon.

%%%%%%%%%%%%%%%%%%%%%%%%%%%%%%%%%%%%
%%%%%%%%%%%%%%%%%%%%%%%%%%%%%%%%%%%

%%%%%%%%%%%%%%%%%%%%%%%%%%%%%%%%%%%%
\begin{figure}[b!]
\vspace*{-8mm}
\bce
\includegraphics[width=.5\textwidth,keepaspectratio,angle=90]{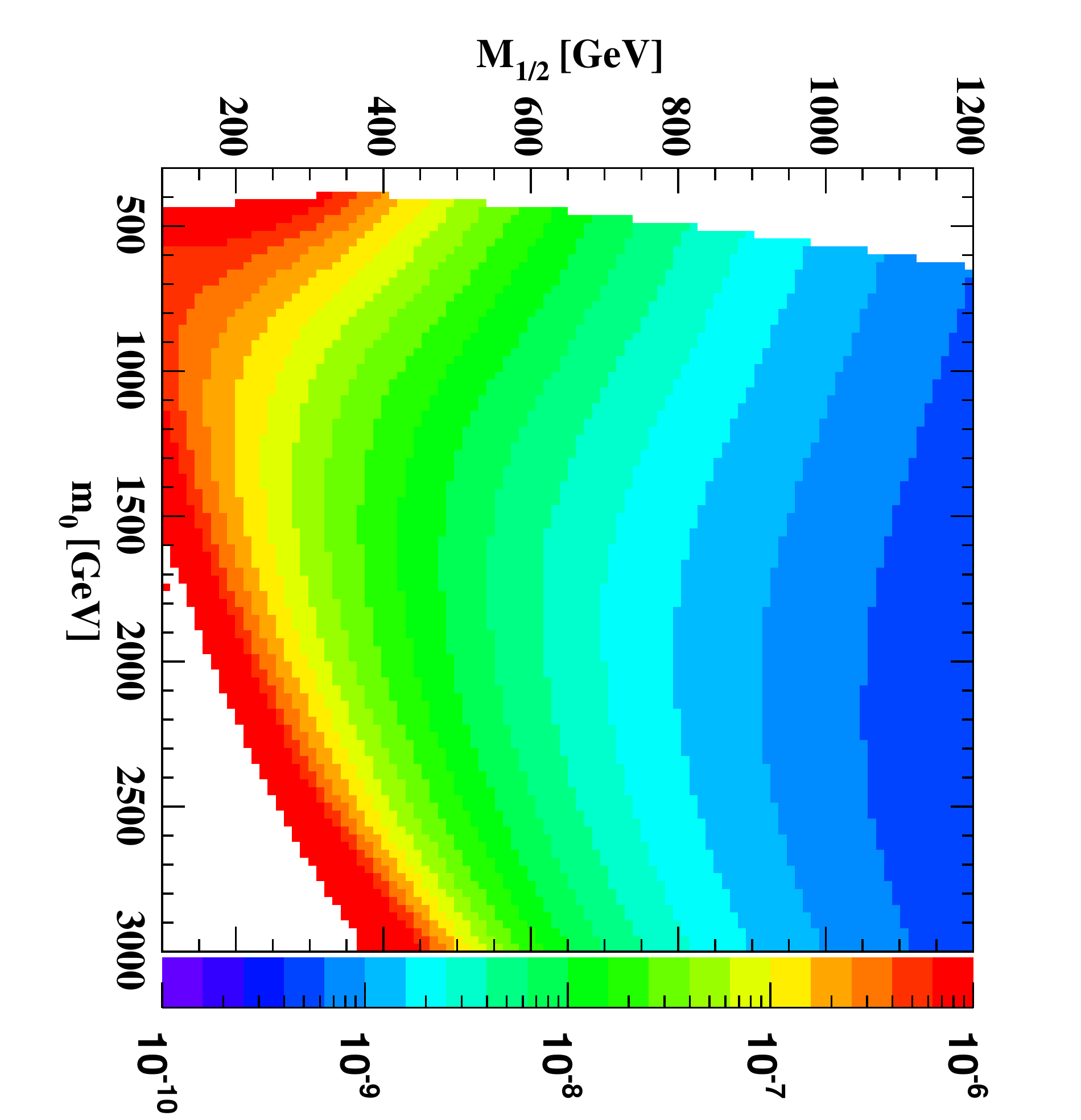}
\ece
\vspace*{-1cm}
\caption{\small Contours of $\ssi$ (in pb) in the $\m$, $\mhf$ plane for $\l=.1$, $\tb=50$, $\A=900$~GeV and $\Ak=60$~GeV (contours for $\Ak=240$~GeV are identical).}
\vspace*{-2mm}
\label{fig:direct_tb50+}
\end{figure}
%%%%%%%%%%%%%%%%%%%%%%%%%%%%%%%%%%%%%

Larger cross sections can be found in the bino-stau coannihilation region for different choices of $\A$, since the mass of the heavy Higgs doublet $H$ as well as its coupling to the LSP depend on $\A$. Contours for the same parameter choice as in sec.~\ref{sec:res}, $\l=.1$, $\tb=50$, $\A=900$~GeV and $\Ak=60$~GeV, are displayed in fig.~\ref{fig:direct_tb50+}. In the bino-stau coannihilation allowed region (small $\m$, $700$~GeV$\lsim\mhf\lsim 1.2$~TeV, see fig.~\ref{fig:tb50+}) $\ssi$ is now in the range $0.6-6\times 10^{-9}$~pb, i.e. within the future experimental reach. The pseudoscalar singlet funnel ($\mhf \sim 200$~GeV, see fig.~\ref{fig:tb50+}) now corresponds to $\ssi\gsim 2\times 10^{-7}$~pb which is already excluded. The value of $\Ak$ has no influence on the SI cross section since the dominant process is the heavy Higgs doublet $H$ exchange and $\Ak$ influences only the singlet Higgs $a/s$ masses. Contours for $\Ak=240$~GeV are therefore basically identical to those for $\Ak=60$~GeV. In this case however, the pseudoscalar singlet funnel ($\mhf \sim 700$~GeV, see fig.~\ref{fig:tb50+}) corresponds to $\ssi\gsim 3\times 10^{-9}$~pb which should be probed by the ton-scale detectors. Finally, as for the $\Ak < 0$ scenarios, most of the focus point region is already excluded by DM direct detection.

%=======================================================================
\section{Higgs searches at the LHC}\label{sec:Higgs}
%=======================================================================

As mentioned in the introduction, the discovery of three neutral Higgs states with large mass splittings would provide a strong evidence in favour of the NMSSM. Furthermore a determination of the masses of the neutral Higgs states, in addition to the measurement of the LSP mass and couplings, is important for a precise theoretical prediction of the DM relic density and/or direct detection cross section. Here we do not address the issue of how precisely masses of sparticles and Higgs bosons can be determined experimentally, however as a first step we explore the discovery potential for the neutral Higgs states of the NMSSM at the LHC. We concentrate on the regions in the parameter space allowed by all phenomenological constraints, which are: the $h$, $a$ and $H/A$ funnels and the stau coannihilation region (the focus point region is disfavoured by the $\gmu$ and DM direct detection constraints).

The most relevant modes for detecting the NMSSM neutral Higgs bosons at the LHC are those that have been studied for the SM and the MSSM. These are: \\[1mm]
\indent 1) associated $b\bar{b}\phi$ production with $\phi \to \tau^+\tau^-$ \\[1mm]
\indent 2) $gg \to \phi \to \gamma\gamma$ \\[1mm]
\indent 3) $gg \to \phi \to ZZ^{(*)} \to 4$ leptons \\[1mm]
\indent 4) $gg \to \phi \to WW^{(*)} \to \ell^+\ell^-\nu\bar{\nu}$ \\[1mm]
\indent 5) $WW \to \phi \to \tau^+\tau^-$ \\[1mm]
\indent 6) $WW \to \phi \to WW$ \\[1mm]
\indent 7) $WW \to \phi \to \gamma\gamma$ \\[1mm]
\noi where $\ell=e,\mu$ and $\phi$ stands for any neutral Higgs scalar or pseudoscalar state, the latter only for channels 1) and 2) though, as a pseudoscalar state does not couple to $WW$ or $ZZ$ at tree level. To compute the statistical significances ($N_\sigma=S/\sqrt{B}$) of all Higgs states in each channel, we have followed the same procedure as in ref.~\cite{Ellwanger:2005uu}. For channels 2-7) this consists simply in multiplying the expected $N_\sigma$ in the SM by the reduced coupling of the considered NMSSM Higgs state to leptons or gauge bosons as compared to those of the SM Higgs with the same mass. For channel 1) the issue is more complicated as the experimental reach is usually given as $5\sigma$ contours in the $m_A$, $\tb$ plane. In the MSSM, these contours correspond to associated production of the heavy doublet states $H/A$ in equal proportion as: ($i$) $m_A \sim m_H$ within the $\tau^+\tau^-$ mass resolution ($ii$) $BR(A\to\tau^+\tau^-) \sim BR(H\to\tau^+\tau^-) \sim 10\%$ in the considered mass range and ($iii$) the $b\bar{b}A$ and $b\bar{b}H$ couplings are equal and scale as $\tb$. The last point explains why this channel is especially important in the MSSM at high $\tb$ but not considered for the SM Higgs boson. In addition, experimental cuts designed to single out the associated production (mainly $b$-tagging) allow us to neglect the contribution of $gg \to H/A$ to the signal. As a result, the net signal rate along the $5\sigma$ contours is twice that for $b\bar{b}H$ or $b\bar{b}A$ alone. Thus, $N_\sigma = 2.5$ would be achieved for $b\bar{b}H$ or $b\bar{b}A$ along these contours were $m_A$ and $m_H$ widely separated. Defining the excluded value of $\tb \equiv t$ as a function of $m_A$ along the contour plots, we compute
\beq
N_\sigma = 2.5 \left( \frac{b_\phi}{t} \right)^2 \frac{BR(\phi \to \tau^+\tau^-)}{0.1}
\eeq
where $\phi$ stands for any NMSSM neutral Higgs scalar or pseudoscalar state and $b_\phi$ is its reduced coupling to $b$ quarks as compared to that of the SM Higgs with the same mass. Finally, in each channel we have combined the signals for degenerate Higgs states with a mass resolution of $15\%$ for channel 1), $1\%$ for channels 2-4) and $10\%$ for channels 5-7). In ref.~\cite{Ellwanger:2005uu} high luminosity (${\cal L} = 300\ \rm{fb}^{-1}$) optimistic results from CMS and ATLAS were used. Here, we take the updated and more realistic low luminosity (${\cal L} = 30\ \rm{fb}^{-1}$) predictions from CMS, summarized in fig.~11.35 for channel 1) and fig.~10.39 for channels 2-7) of the CMS-TDR~\cite{Ball:2007zza}.

In order to explore the LHC reach for NMSSM Higgs searches, we have plotted contours of the maximal significance for each Higgs state in the plane $\m$, $\mhf$ with the same parameter choices as in sec.~\ref{sec:res}-\ref{sec:dd}. Rather than showing a large number of plots, we will now discuss some general properties and eventually present a few selected case studies. First, let us remark that for all parameter choices, the maximal significance for the lightest Higgs doublet $h$ is always $N_\sigma(h) \sim 4$, but it hardly reaches 5 in a single channel. This situation (similar to that of the MSSM) means that a high luminosity run, improved detection techniques or channel combinations will be necessary in order to exhibit a clean $5\sigma$ signal for the light $h$ at the LHC. The best channels for $h$ discovery are either 2)~$gg \to h \to \gamma\gamma$ or 5)~$WW \to h \to \tau^+\tau^-$. However, other channels as 3)~Ê$gg \to h \to ZZ^{(*)} \to 4$ leptons and 7)~Ê$WW \to h \to \gamma\gamma$ can yield a significance $\sim 1-3$ (we have not combined significances in different channels for each Higgs state). The heavy doublet states $H/A$ are always nearly degenerate and the LHC will not be able to see them separately. The best channel for these states is 1) associated $b\bar{b}H/A$ production with $H/A \to \tau^+\tau^-$. This channel covers masses up to $m_{H/A} \sim 800$~GeV and is enhanced by a factor $\tan^2\!\beta$. Hence, a $5\sigma$ signal shall be seen at the LHC provided that $\mhf$ is not too large and $\tb$ is large. In our selected scans, we indeed found that for $\tb = 5$ or $10$ the significances for $H/A$ are always small and $N_\sigma(H) = N_\sigma(A) = 5$ is reached only for $\tb = 50$ and $\mhf$ not too large.

In the singlet sector, we found that the scalar $s$ is usually too heavy and/or too decoupled to be seen at the LHC. This is because if the singlet $s$ is light and $\l$ is not too small ({\it ie} the doublet/singlet mixings are substantial) then mixing effects between $h$ and $s$ always lead to a light state in the Higgs spectrum, already excluded by LEP constraints (note that this is not the case for a light pseudoscalar singlet $a$ which does not mix with the light $h$). On the other hand, if the singlet $s$ is light and $\l$ is small ({\it ie} the doublet/singlet mixings are negligible) then one obtains an effective MSSM with a decoupled singlet sector. One can still have differences between such a scenario and the true MSSM if the singlino is the LSP (cf. sec.~\ref{sec:sing}). However, the couplings of the singlet Higgs states to matter fermions and gauge bosons are then so small that it would be impossible to see them at the LHC (we will not discuss here the possibility of observing a singlino LSP at the LHC). The significances for the scalar singlet $s$ are therefore always small. The only exception is when it is nearly degenerate with the heavy doublet states $H/A$. In this case, the significances for the three states $H$, $A$ and $s$ are equal and can be large, although the net contribution of the singlet $s$ to the total significance is always small and the LHC would see only one peak.

The pseudoscalar singlet $a$ can play an important role in DM relic density computation, as we have seen in sec.~\ref{sec:pseudo}. Contrary to the scalar singlet $s$ it can be relatively light without being excluded by LEP. If it is not too decoupled ({\it ie} if $\l$ is not too small) one can then expect to be able to see it at the LHC. For the selected scans of sec.~\ref{sec:res}-\ref{sec:dd}, we have checked that for $\l=.01$ the significances for the singlet $a$ are always small. For $\l=.1$ we have found some cases where these significances are large enough to be able see something at the LHC. These cases correspond to $\tb = 50$ only as the main discovery channel for the singlet $a$ is 1)  associated $b\bar{b}a$ production with $a \to \tau^+\tau^-$, which is $\tb$ enhanced. We will now study in details a few cases, summarized in table~\ref{tab:lhcsig}.

Let us start with $\A=-900$~GeV, $\Ak=-60$~GeV as in fig.~\ref{fig:tb50-}(a). After imposing all phenomenological constraints, the only region allowed is the $a$ funnel for $\m\sim2$~TeV and $\mhf\sim200$~GeV. Case (1) in table~\ref{tab:lhcsig} corresponds to a point in this region. The significance for the pseudoscalar singlet in this case is very small ($N_\sigma(a) = .1$). Here, the mass of the scalar singlet is $m_s = 567$~GeV, close enough to $m_H$, $m_A$ to be within the $\tau^+\tau^-$ resolution. Hence $N_\sigma(s) = N_\sigma(H) = N_\sigma(A) = 13$, although the LHC would see only one peak. The significance for the light doublet $N_\sigma(gg \to h \to \gamma\gamma)$ is below 5. However, $N_\sigma(gg \to h \to ZZ^{(*)} \to 4\ \rm{leptons}) = 2.7$, $N_\sigma(WW \to h \to \tau^+\tau^-) = 4.0$ and $N_\sigma(WW \to h \to \gamma\gamma) = 2.2$, so that combining the signals from different channels would yield a global significance above 5. The situation for the light $h$ is similar in all our case studies.

%%%%%%%%%%%%%%%%%%%%%%%%%%%%%%%%%%%%
\begin{table}[t!]
\footnotesize
\begin{center}
\begin{tabular} {|c|r|r|r|r|r|r|r|c|r|r|c|r|c|}
\hline
Case & $\A$ & $\Ak$ & $\m$ & $\mhf$ &$\Omega h^2$ & $m_{\chi^0_1}$ & $m_h$ & $N_\sigma(h)$ & $m_H$  & $m_A$ & $N_\sigma(H/A)$ & $m_a$ & $N_\sigma(a)$ \\ \hline
(1) &  -900  & -60 & 1900 & 207 & .129 &   87 & 114 & 4.1 (2) & 528 & 527 & 13 (1) & 183 & 0.1 (1) \\
(2) & -1500 & -60 & 1600 & 130 & .133 &   56 & 115 & 4.2 (2) & 434 & 441 & 17 (1) & 191 & 1.0 (1) \\
(3) &   900  &   60 & 1625 & 155 & .130 &   61 & 114 & 4.4 (5) & 296 & 309 & 26 (1) & 190 & 3.9 (1) \\
(4) &   900  &   60 & 1630 & 305 & .035 & 123 & 114 & 4.4 (5) & 315 & 298 & 27 (1) & 200 & 5.9 (1) \\
(5) &   900  &   60 &   900 & 650 & .133 & 273 & 114 & 4.2 (5) & 476 & 482 & 19 (1) & 251 & 0.8 (1)  \\
(6) &   900  &   60 &   475 & 740 & .110 & 311 & 114 & 4.2 (5) & 420 & 430 & 23 (1) & 253 & 1.7 (1)  \\
\hline
\end{tabular}
\end{center}
\caption{Higgs masses and statistical significances at the LHC with ${\cal L}=30fb^{-1}$ for selected points. In all cases $\tb=50$, $\l=.1$ and $m_t=172.6$~GeV (except case 4 where $m_t=171.2$~GeV). The channel where the significance is maximum is indicated in parenthesis. Dimensionful parameters are in GeV.}
\vspace*{-2mm}
\label{tab:lhcsig}
\end{table}
%%%%%%%%%%%%%%%%%%%%%%%%%%%%%%%%%%%%

For $\A=-1500$~GeV, $\Ak=-60$~GeV as in fig.~\ref{fig:tb50-}(b), there are 3 regions allowed by all phenomenological constraints. In the $a$ funnel, as well as in the stau coannihilation region, one finds very small significances for the singlet $a$, $N_\sigma(a) \sim .1$. However, in the $h$ funnel one finds slightly larger significances. This is illustrated by case (2) in table~\ref{tab:lhcsig} where $N_\sigma(a) = 1$. Higher luminosity or improved detection techniques could eventually lead to a $5\sigma$ signal for the pseudoscalar singlet in this scenario.

Next, we consider the case where $\A=900$~GeV, $\Ak=60$~GeV as in fig.~\ref{fig:tb50+}(a). Here, the allowed regions are the $h$, $a$ and $H/A$ funnels, the stau coannihilation region and the focus point region (although the latter is usually disfavoured by the $\gmu$ and DM direct detection constraints). Case (3) in table~\ref{tab:lhcsig} corresponds to a point in the $h$ funnel where all significances are quite large, making it possible to distinguish the NMSSM from the MSSM. Even larger significances for the singlet $a$ ($N_\sigma(a) > 5$) are found at smaller values of $\mhf$, although these are disfavoured by the $\btau$ constraints. Note that with a smaller value for the top quark mass the doublet/singlet ($A/a$) mixing is increased and the significance for the singlet $a$ is already above 5 as can be seen in case (4) of table~\ref{tab:lhcsig}. This point passes all experimental constraints. However, the bino annihilation cross section is large due to the singlet $a$ resonance and the DM relic density is below the range allowed by cosmological data. In such a scenario, one would then have to rely on a different mechanism to explain the abundance of DM in the universe. Finally, case (5) sits in the $H/A$ funnel and case (6) in the stau coannihilation band. In both cases, we find substantial significances for the singlet $a$, although always below $5\sigma$. As for case (2), a higher luminosity or improved detection techniques could eventually lead to a $5\sigma$ signal for the singlet $a$ in these scenarios. Larger values of $\Ak$, eg $\Ak = 240$~GeV as in fig.~\ref{fig:tb50+}(a), yield a heavier singlet $a$ and therefore smaller significances.

%=======================================================================
\section{Conclusion}\label{sec:dis}
%=======================================================================

We have reexamined the DM favoured scenarios in the CNMSSM taking into account constraints from B-physics observables and from the $\gmu$. Although the allowed parameter space is strongly constrained by these observables we confirm our previous results : scenarios where the LSP annihilates near a pseudoscalar Higgs singlet resonance or where the LSP is mainly singlino are allowed in addition to the MSSM-like scenarios such as bino-sfermion coannihilation, bino annihilation through a Higgs doublet resonance or mixed higgsino LSP.

B-physics observables constrain mainly the scenarios at large $\tb$ with $\bsg$ being more effective in the low $\m$, $\mhf$ region when $\A$ is negative while $\bsmu$ and $\btau$ being more powerful constraints when $\A>0$ and/or at large $\m$. The most powerful constraints on the parameter space are however obtained from the $\gmu$. Indeed, explaining the deviation from the SM by \susy\ contributions requires light sfermions and charginos, especially at low values of $\tb$. The focus point region at large $\m$ as well as the bino-sfermion coannihilation region at large $\mhf$ are therefore strongly constrained. This feature is not specific to the CNMSSM and is also observed in the CMSSM. A significant portion of the Higgs (doublet or singlet) funnel regions is also constrained by $\gmu$.

The recent upper limits on the neutralino proton elastic scattering cross section constrain mainly the focus point region at large $\tb$. When the LSP is partially higgsino and $\tb$ is large, the predictions from $\ssi$ rise very fast with the decrease of the mass of the heavy Higgs doublet so one will easily probe all the allowed parameter space in the near future. This is so even when taking into account the large uncertainties in the theoretical predictions. In fact we have shown that future direct detection experiments have good prospects for probing the parameter space of the CNMSSM with a dominantly bino or higgsino LSP. On the other hand, at small values of $\tb$, or when the Higgs and squarks are in the TeV range, one will have to wait for the large scale detectors. This statement however is strongly correlated with the lower limit on $\amu$ we have imposed which  disfavours the scenarios at large values of $\mhf$ with little prospects for direct detection. The large theoretical uncertainties in the computation of the direct detection rate is an issue. In particular a better determination of the quark coefficient in the nucleon could significantly reduce the uncertainty. Finally the characteristic NMSSM scenario with a singlino LSP is far beyond the reach of large scale detectors. Similar conclusions were reached in an analysis performed in the context of another constrained version of the NMSSM
where $\Ak=A_0$~\cite{Balazs:2008cp}.

In the near future with the onset of the LHC, improved sensitivity on B-physics observables and improved direct detection results, prospects for discovery of supersymmetry are good. If a clean signal of physics beyond the standard model is observed, the issue will be to establish whether or not the CNMSSM (or even only the NMSSM) is the correct scenario. What distinguishes the model from the MSSM is the more elaborate neutralino and Higgs sectors. We have analysed the potential of the LHC to unambiguously identify the NMSSM by observing at least 3 neutral Higgs states. We have found that this is possible only for large values of $\tb \sim 50$ and $\l \sim .1$ where one can benefit both from the enhanced couplings of $H/A$ to $b\bar{b}$ and $\tau^+\tau^-$ and from substantial doublet/singlet mixings. The best signals are found when the heavy doublets $H/A$ and the pseudoscalar singlet $a$ are at most a few hundred GeV's (that is at low $\mhf$ or $\m$). This region is also the one that will be efficiently probed by direct detection searches. At lower values of $\tb$, there is little prospects of discovery at the LHC for the Higgs singlet states. This means that in the scenarios where the pseudoscalar singlet $a$ contributes significantly to the bino annihilation, the only indication in favour of the NMSSM would be that the MSSM prediction for the DM relic density is much above the cosmological measurements.

Note also that a very light pseudoscalar singlet $a$ can appear in the NMSSM and it is not completely excluded by LEP constraints nor by recent $\Upsilon$ radiative decay constraints~\cite{Dermisek:2006py,SanchisLozano:2007wv,Domingo:2007dx,Domingo:2008rr}. It could even explain the small $2.3\sigma$ excess in the $e^+e^- \to Z b\bar{b}$ channel at LEP if the light doublet $h$ decays mainly as $h \to aa$~\cite{Dermisek:2005ar,Dermisek:2005gg,Dermisek:2006py,Dermisek:2007yt}. In this case, new detection channels might reveal necessary in order to reinstate the 'no-lose' theorem for NMSSM Higgs discovery at the LHC\cite{Ellwanger:2005uu,Barger:2006sk,Moretti:2006hq,Forshaw:2007ra,Carena:2007jk,Cheung:2007sva,Belyaev:2008gj,Djouadi:2008uw}. Finally, the observation of a singlino at the LHC or at the ILC would of course be another method to provide evidence for the NMSSM~\cite{Ellwanger:1997jj,MoortgatPick:2005vs,Kraml:2005nx,Barger:2006kt,LH:2008gva,Kraml:2008zr}.

%=======================================================================
\section*{Acknowledgments}
%=======================================================================

We thank U. Ellwanger and F. Domingo for helpful discussions. We are grateful 
to A. Nikitenko for providing us with the tables for CMS Higgs studies. 
This work was supported in part by the GDRI-ACPP of CNRS. The work of 
A.P. was supported by the Russian foundation for Basic Research, 
grant RFBR-08-02-00856-a and RFBR-08-02-92499-a.

\section*{Note added}
A recent estimate of the hadronic contribution to $\gmu$ based on  the latest 
 data from BaBar indicates that $\amu$, the deviation between the measured value and the theoretical SM prediction,
  may be smaller than considered here ~\cite{davier}. If this result is confirmed, the $\amu$ constraint in the large $\mhf$ region would disappear.

%=======================================================================
\providecommand{\href}[2]{#2}\begingroup\raggedright\endgroup

%=======================================================================

\end{document}